\documentclass[aps,prc,reprint,superscriptaddress,amsmath,amssymb]{revtex4-1}

\usepackage{setspace}
\usepackage{macros}
\usepackage{graphicx}
\usepackage{bm}
\usepackage{hyperref}
\begin{document}


\title{Direct measurement of the low energy resonances in $^{22}\rm{Ne}(\alpha,\gamma)^{26}\rm{Mg}$ reaction}



\author{Shahina}
\author{J.~G\"orres}
\author{D.~Robertson}
\author{M.~Couder}
\author{O.~Gomez}
\author{A.~Gula}

\affiliation{Department of Physics and Astronomy, University of Notre Dame, Notre Dame, Indiana 46556, USA}
\affiliation{The Joint Institution of Nuclear Astrophysics-Center for the Evolution of the Elements, University of Notre Dame, Notre Dame, Indiana 46556, USA}

\author{M.~Hanhardt}
\affiliation{Department of Physics, South Dakota School of Mines and Technology, Rapid City, South Dakota 57701, USA}
\affiliation{South Dakota Science and Technology Authority, Sanford Underground Research Facility, Lead, South Dakota 57754, USA}
\author{T.~Kadlecek}
\affiliation{Department of Physics, South Dakota School of Mines and Technology, Rapid City, South Dakota 57701, USA}

\author{R.~Kelmar}

\author{P.~Scholz}
\author{A.~Simon}
\author{E.~Stech}
\affiliation{Department of Physics and Astronomy, University of Notre Dame, Notre Dame, Indiana 46556, USA}
\affiliation{The Joint Institution of Nuclear Astrophysics-Center for the Evolution of the Elements, University of Notre Dame, Notre Dame, Indiana 46556, USA}

\author{F.~Strieder}
\affiliation{Department of Physics, South Dakota School of Mines and Technology, Rapid City, South Dakota 57701, USA}

\author{M.~Wiescher}
\affiliation{Department of Physics and Astronomy, University of Notre Dame, Notre Dame, Indiana 46556, USA}
\affiliation{The Joint Institution of Nuclear Astrophysics-Center for the Evolution of the Elements, University of Notre Dame, Notre Dame, Indiana 46556, USA}

\date{\today}

\begin{abstract}
 The $^{22}\rm{Ne}(\alpha,\gamma)^{26}\rm{Mg}$ is an important reaction in stellar helium burning environments as it competes directly with one of the main neutron sources for the s-process, the $^{22}\rm{Ne}(\alpha,n)^{25}\rm{Mg}$ reaction. The reaction rate of the $^{22}\rm{Ne}(\alpha,\gamma)^{26}\rm{Mg}$ is dominated by the low energy resonances at $E_{\alpha}^{lab}$ = 650 and 830 keV respectively. The $E_{\alpha}^{lab}$ = 830 keV resonance has been measured previously, but there are some uncertainties in the previous measurements. We confirmed the measurement of the $E_{\alpha}^{lab}$ = 830 keV resonance using implanted $^{22}$Ne targets. We obtained a resonance strength of $\omega\gamma$ = 35 $\pm$ 4 $\mu eV$, and provide a weighted average of the present and previous measurements of $\omega\gamma$ = 35 $\pm$ 2 $\mu eV$ with reduced uncertainties compared to previous studies. We also attempted to measure the strength of the predicted resonance at $E_{\alpha}^{lab}$ = 650 keV directly for the first time and found an upper limit of $\omega\gamma$  $\mathrm{<0.15}$ $\mu eV$ for the strength of this resonance. In addition, we also studied the $E_{P}^{lab}$= 851 keV resonance in $^{22}\rm{Ne}(p,\gamma)^{23}\rm{Na}$, and obtained a resonance strength of $\omega\gamma$ = 9.2 $\pm$ 0.7 eV with significantly lower uncertainties compared to previous measurements. 
 
\end{abstract}


\maketitle


\section{Introduction}

The slow neutron capture process (s-process) is responsible for the creation of half the elements beyond iron. It takes place when the timescale between successive neutron captures is much larger than their lifetime of $\beta$-decay ~\cite{Rolfs.etal.89}. Hence, less neutron rich nuclei are produced along the line of $\beta$ stability. The s-process can be categorized into two types: the weak s-process responsible for the creation of elements over the range 60 $<$ A $<$ 90, and the main s-process which creates nuclei with A $>$ 90~\cite{bisterzo_branchings_2015}. The temperature range of interest is 0.2 to 0.6 GK for both scenarios. Both processes require a sufficiently large neutron flux. However, the source of these neutrons is still under debate. In AGB stars, the main s-process nucleosynthesis occurs during the late stages of stellar evolution when the star has a degenerate C-O core, a thin radiative He-intershell and an expanded convective envelope. During the AGB phase, the star experiences a series of flashes called thermal pulses which are triggered by the sudden activation of the 3$\alpha$-process at the base of the He-intershell at highly degenerate gas conditions~\cite{bisterzo_branchings_2015}. In this intershell environment, two important reactions are activated, $\Can$ prior to the thermal pulse and $\Nean$ during the thermal pulse, which act as the main sources of free neutrons for the s-process. Both $\Nean$ and $\Can$ provide the neutron flux in low mass-AGB stars, $\Can$ being the dominant contributor. The $^{13}$C buildup occurs in a thin zone (the $^{13}$C pocket) via the capture of mixed-in hydrogen on the abundant $^{12}$C via $^{12}$C(p,$\gamma$)$^{13}$N($\beta^+\nu$)$^{13}$C. $\Can$ burns radiatively for an extended time releasing a high neutron flux, when the temperature exceeds 0.8 $\times 10^{8}$ K~\cite{bisterzo_branchings_2015}. The second dominant neutron source, $\Nean$, has a negative Q-value of -478 keV, and thus requires a sufficiently high temperature. Such conditions are reached only during the advanced thermal pulses in low mass AGB stars and the core helium burning phase of massive stars. 

$\Nean$ also acts as the main neutron source for the weak s-process in core helium burning of massive stars~\cite{Kaeppeler}. The $^{22}$Ne is produced from the large amount of $^{14}$N left at the end of the CNO cycle in the H-burning core via the following reaction chain: 
\begin{align}
^{14}\rmN(\alpha,\gamma)^{18}\rmF(\beta^+,\nu)^{18}\rmO(\alpha,\gamma)^{22}\rm{Ne}
\end{align}
However, because of its negative Q-value the temperature during core helium burning is not sufficient to trigger the ignition of $\Nean$ at once. Only towards the end of the helium burning phase with the stellar core gradually contracting, high temperature and density conditions sufficient for the above reaction are reached. Hence, the $\Nean$ reaction is operative only during the peak of the helium flash and at the end of core helium burning. 

At lower temperature the neutron-producing role of $\Nean$ is complicated by the competing $\Neag$ reaction. The $\Neag$ reaction has a positive Q-value $=10614.74\pm 0.03$ keV and, therefore, starts operating at relatively low temperature, before $\Nean$ can kick in. As a consequence, it is important to investigate the reaction rate of $\Neag$ in order to constrain the neutron production for the weak s-process. 
 
 At stellar helium burning temperature, low energy resonances at $E_{\alpha}^{lab}$ = 650 and 830 keV lie within the Gamow window and dominate the $\Neag$ reaction rate. $\Neag$ proceeds through resonant capture reaction mechanism, thus populating natural parity states in the compound nucleus $^{26}$Mg.\cite{Talwar.etal2016} Above T=0.3 GK, the rate is dominated by the $E_{\alpha}^{lab}$ = 830 keV resonance which has been observed in both $(\alpha,n)$ and $(\alpha,\gamma)$ direct measurement experiments. 
 
 The $\Neag$ reaction was  first measured directly by Wolke \textit{et al.}~\cite{Wolke.etal1989} for $E_{\alpha}^{lab}$ = 0.71 to 2.25 MeV. They found 15 new resonances in the energy range covered, with the lowest energy resonance at 830 keV. They concluded that the reaction rate is strongly impacted by the 830 keV resonance, whose resonance strength was measured to be $\strength$ = $36 \pm 4$ $\mu$eV. For this measurement, they used a differentially pumped gas target system with Neon gas enriched to $99\% $ in $^{22}\rm{Ne}$. The $\gamma$ ray transitions were detected with Ge(Li) detectors. A recent direct measurement of this resonance was also performed by Hunt \textit{et al.}~\cite{Hunt.etal.19} who obtained its resonance strength to be $\strength$ = $46 \pm 12$ $\mu$eV. Unlike Wolke \textit{et al.}~\cite{Wolke.etal1989}, they used implanted $^{22}\rm{Ne}$ targets to determine the resonance strength from the thick target yield curve. They detected the entire cascade of $\gamma$ rays using a $\gamma\gamma$-coincidence spectrometer, consisting of high-purity germanium (HPGe) detector surrounded by a NaI(Tl) annulus.
 
Complementary to the direct measurements, Talwar \textit{et al.}~\cite{Talwar.etal2016} performed indirect measurements using ($^{6}$Li,d) $\alpha$- transfer and $(\alpha,\alpha^{'})$ inelastic scattering techniques to populate states that are most likely to appear as resonances in the $\Neag$ reaction. They observed a  strong transition in both ($^{6}$Li,d) and $(\alpha,\alpha^{'})$ measurements at $\rmE_{x}=11.167$ MeV, which corresponds to a resonance in the $^{22}$Ne + $\alpha$ system at $E_{\alpha}^{lab}$ = 650 keV. They assigned a spin-parity of $J^{\pi}= (1^{-},2^{+})$ to this state, and obtained a spectroscopic factor of $S_{\alpha}=0.36$ (corresponding to $\Gamma_{\alpha}=0.18$ $\mu$eV). Based on this, they suggested a resonance strength of $\omega\gamma =$ 0.54 $\pm$ 0.07 $\mu$eV. They concluded that this state completely dominates the $\Neag$ reaction rate between T $\simeq 0.2-0.4$ GK and would also potentially dominate the $\Nean$ reaction rate below 0.2 GK. Therefore, it is essential to measure this low energy resonance in $\Neag$ in order to accurately predict the total neutron flux for the weak s-process. 

In this work, we present the direct measurements of these two low-energy resonances at $E_{\alpha}^{lab}$ = 650 and 830 keV which play an important role in the reaction rate of $\Neag$. Several resonances are expected below the 650 keV resonance \cite{Talwar.etal2016}. However, the penetrability rapidly drops with beam energy and the corresponding alpha widths are at least two or three orders of magnitude lower. Hence, the resulting resonance strength would be well below the sensitivity of this experiment. For this reason, we concentrated on the search for the 650 keV resonance which would have the largest impact on the neutron production for the weak s-process if the predicted strength is confirmed \cite{Talwar.etal2016}.

\section{Experimental setup} \label{sec::exp_setup}

\subsection{Target Preparation and Characterization} \label{subsec::target}

The $^{22}$Ne targets were made and tested at the University of Notre Dame using the 5U St.Ana Pelletron accelerator. $\mathrm{Ne}$ gas enriched in $\Ne$ to 99$\%$ was used to produce a $^{22}$Ne beam with an energy of 200 keV, which was implanted onto thick Ta backings, 3.8 cm $ \times$ 3.8 cm in size. Tantalum was used as a backing, as it has a high concentration of $^{22}$Ne at implantation saturation~\cite{SELIN1967218} and beam induced reactions with the backing are minimized. The $^{22}$Ne beam shape was defined by slits in front of the target, and the beam was wobbled in both x and y directions to have uniform implantation of $^{22}$Ne onto Ta. To measure the saturation curve several $^{20}$Ne targets with increasing implantation dosage of 30, 60, 90, 120, 150, 180,
225, 270 and 300 mC were produced. The saturation was achieved near 300 mC, and hence all the implanted $^{22}$Ne targets used during the experiment had a 300 mC dosage over a target area of 3 cm$^2$. 

\begin{figure*}
    \centering
    \includegraphics[width=0.9\textwidth]{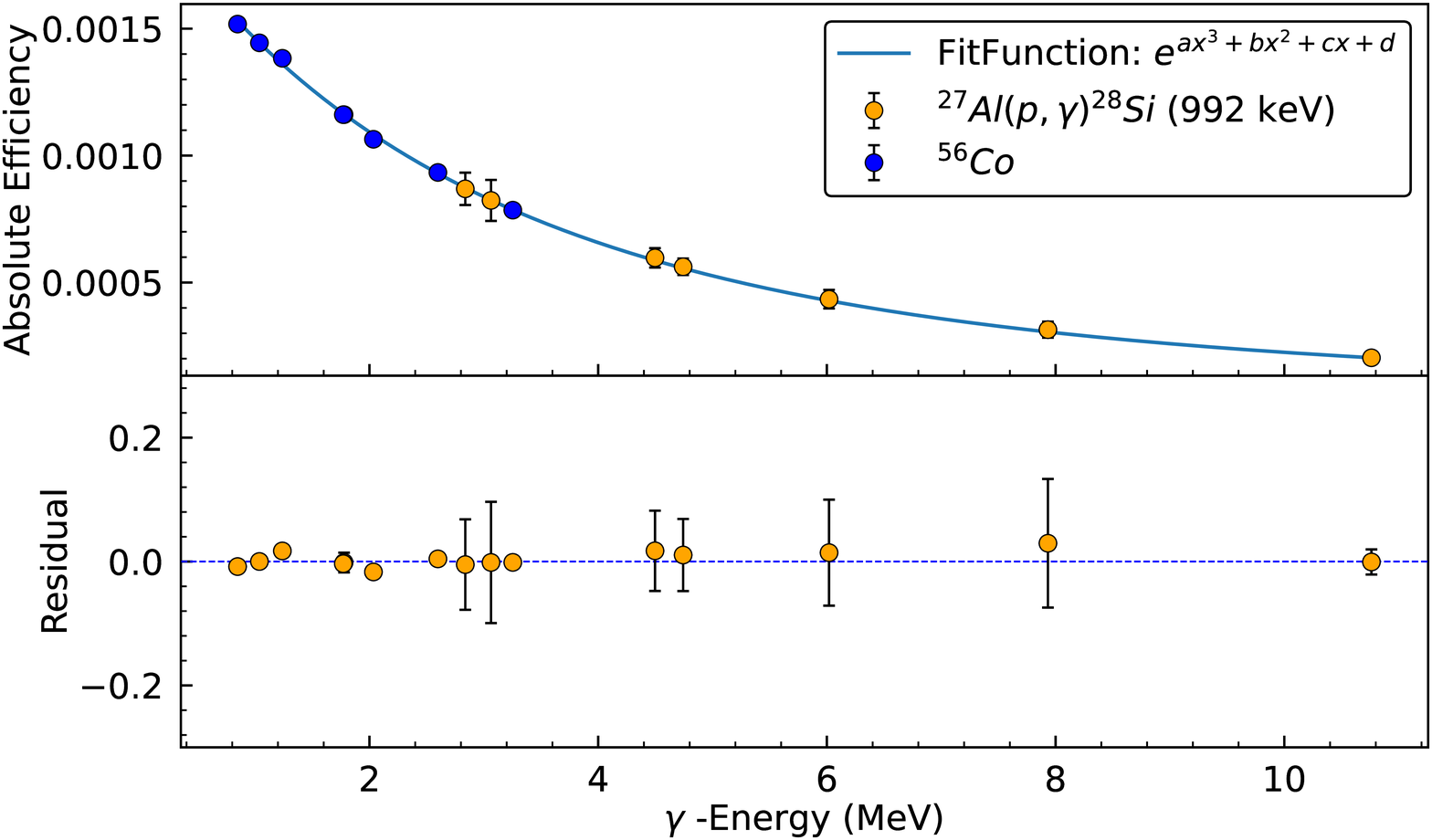}
    \caption{Absolute $\gamma$-ray detection efficiency as a function of energy for the Ge detector used for the determination of target stoichiometry.}
    \label{efficiency_curve}
\end{figure*}

The implantation profile of the $^{22}$Ne targets was investigated using the 1278 keV resonance in the $\Nepg$ reaction~\cite{Depalo}. The efficiency of the emitted $\gamma$-rays was measured by using a p-type coaxial germanium detector operating at +4000V. The detector was placed at an angle of 55$^{\circ}$ with respect to the beam direction to minimize angular distribution effects. The distance of the detector from the target was 20 cm to reduce summing effects. Moreover, a 1 mm thick lead sheet was attached to the front face of the detector to reduce the $\gamma$-ray flux from inelastic scattering from the Ta backing. 
 The absolute full-energy peak efficiency was measured using calibrated radioactive sources like  $^{56}$Co and $^{60}$Co. The sources were placed exactly at the target position, 20 cm from the detector to have the same geometry. The efficiency curve was extended to 10.76 MeV using the well known resonance at 992 keV in the $\Alpg$ reaction~\cite{Antilla}. The absolute efficiency curve for the Ge detector is shown in Fig.~\ref{efficiency_curve}. It was obtained by normalizing the efficiency at higher energies, to the efficiency of the 1779 keV $\gamma$-ray in the $\Alpg$ resonance, taking into account the well-known branching ratios and angular distributions~\cite{Antilla}. The relative efficiency was converted into absolute efficiency using the yield value of $(1.08 \pm 0.06) \times 10^{-9}$ $\gamma-$ rays (1779 keV) per incident proton~\cite{Antilla}, which corresponds to a resonance strength of 1.93 $\pm$ 0.13 eV~\cite{PAINE1979389}. 

The stoichiometry of the $^{22}$Ne targets was determined using the well-known resonance at  1278 keV in $\Nepg$~\cite{Depalo} using the thick target yield formula~\cite{Iliadis},
\begin{equation}
    \omega \gamma = \frac{2 Y\epsilon_{\mathrm{eff}}}{\lambda^{2}\eta},
    \label{eq:1}
\end{equation}
where $Y$ is the yield, $\eta$ is the efficiency, $\lambda$ is the deBroglie wavelength of the incident particle and $\epsilon_{\mathrm{eff}}$ is the effective stopping power of protons given by: 
\begin{equation}
    \epsilon_{\mathrm{eff}} =\frac{M_{Ne}}{M_{P} +M_{Ne}} \left[\epsilon_{Ne} + \left(\frac{N_{Ta}}{N_{Ne}}\right)\epsilon_{Ta}\right],
    \label{stoi}
\end{equation}
where $N_{Ta}/N_{Ne}$ is the initial target stoichiometry, $M_{P}$ and $M_{Ne}$ are the masses of the projectile and target nuclei (in amu), while $\epsilon_{Ne}$ and $\epsilon_{Ta}$ denote the stopping power of the protons in Ne and Ta respectively, which were obtained using SRIM~\cite{SRIM}. The quoted uncertainties of these values for the relevant proton energy are 1.6$\%$ and 3$\%$ for Ne and Ta, respectively. 

\begin{figure}
  \centering
  \includegraphics[width=0.5\textwidth]{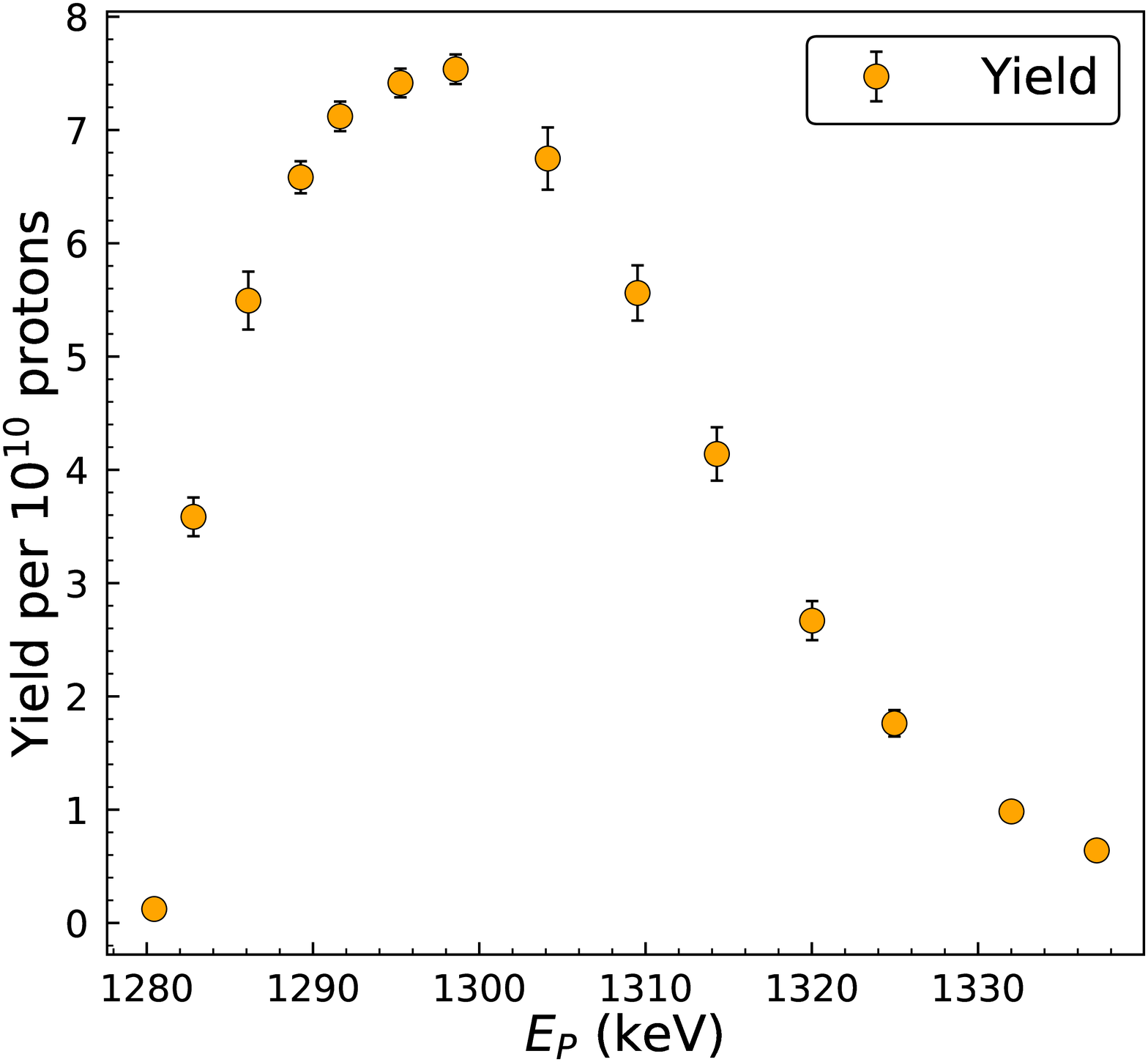}
  \caption{Yield curve for the $E_{P}^{\mathrm{lab}}$= 1278 keV resonance in the $\Nepg$ reaction.}
  \label{1278_RS}
\end{figure}

The yield curve of the 1278 keV resonance was measured with protons from the St.Ana accelerator utilizing the same target station which has been used for the Ne implantation. The yield curve is shown in Fig.~\ref{1278_RS}, which is obtained after taking the average of the absolute yield calculated using the following $\gamma$-ray transitions: 2540, 2980, 3914, 6102, 7036 and 9576 keV. The yield from the individual gamma rays was corrected with the branching ratio obtained from~\cite{BAKKUM19891} and angular distribution obtained from~\cite{viitasalo_angular_1972}. The yield curve in Fig.~\ref{1278_RS} depicts the distribution of the implanted $\Ne$ in the Ta backing. In order to obtain the stoichiometry of the implanted target, Equation.~\ref{eq:1} was first used to calculate the effective stopping power of the incident protons, using the resonance strength of $\mathrm{\omega \gamma}$ = 10.8 $\pm$ 0.7 eV from \cite{Depalo}. The effective stopping power thus obtained was then substituted in Equation.~\ref{stoi} to compute the stoichiometry of the target at maximum yield, which turned out to be $N_{Ta}/N_{Ne}$= $2.7 \pm 0.3$. The total number $n$, of $\mathrm{Ne}$ nuclei implanted onto the Ta backing was thereafter obtained using the formula~\cite{Iliadis},
\begin{equation}
    n = \frac{2A_{Y}}{\lambda^{2} \omega\gamma},
    \label{ne_toms}
\end{equation}
where $\mathrm{A_{Y}}$ is the area under the $\Nepg$ resonance yield curve. $\mathrm{A_{Y}}$ was obtained by numerical integration of the excitation curve shown in Fig.~\ref{1278_RS}. The total number of $\mathrm{Ne}$ nuclei implanted, $n$, was found to be (6.21 $\pm$ 0.37) $\times$ $10^{17}$ $\mathrm{atoms/cm^{2}}$. The error of the integration and the statistical errors are small compared to the uncertainty of the resonance strength. The value obtained for the total number of implanted Ne nuclei is consistent with the implantation dose within uncertainty and indicates an implantation efficiency of nearly 100$\%$. 

\subsection{CASPAR accelerator} \label{subsec::caspar}

\begin{figure}
    \centering
    \includegraphics[width=0.4\textwidth]{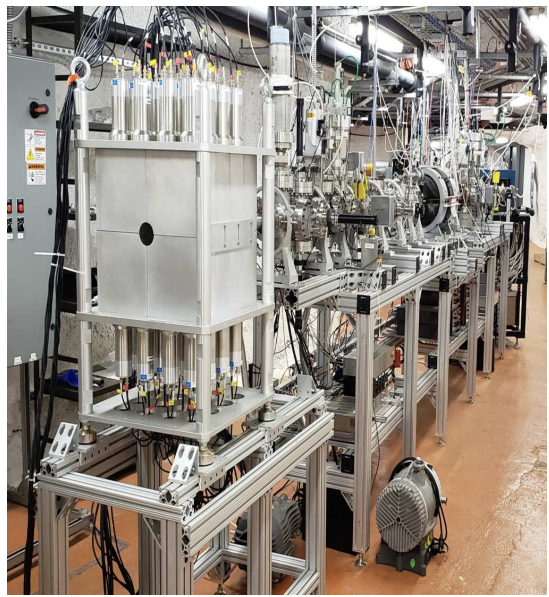}
    \caption{Experimental setup showing HECTOR placed at the end of the beamline at the CASPAR underground lab. The target was placed at the center of the detector cube.}
    \label{HECTOR}
\end{figure}

The measurement of the low energy resonances in $\Neag$ was performed at CASPAR (Compact Accelerator System for Performing Astrophysical Research), located on the 4850 ft. level of the Sanford Underground Research Facility in Lead, South Dakota, at the former site of the Homestake Gold mine~\cite{CASPAR,caspar_site}. The rock above acts as an effective shield from the cosmic-ray induced background, thus helping to measure the weak resonances which have a low signal-to-background ratio. The 1 MV JN positive ion accelerator at CASPAR was used to deliver protons and alpha beams onto the target with intensity ranging from 10 to 100 $\mu$A. The energy calibration of the beam was performed by measuring the front edges of the resonances in $\Oag$ at 750 and 767 keV~\cite{Vogelaar} and scanning the well known resonances in $\Alpg$~\cite{osti_4398955}. The beam energy uncertainty amounted to $\pm$2 keV. The beam was sent through an analyzing magnet which bent it by 25$^{\circ}$ before transporting it to the experimental end station. The beam was defocused and delivered to the target. The size of the beam spot was nearly 1 cm in diameter. The target was mounted inside the High EffiCiency TOtal absorption spectrometeR (HECTOR)~\cite{Craig} at the end of the beamline, as shown in Fig.~\ref{HECTOR}. The setup is identical to the setup used for the measurement of $\Alpg$ resonances \cite{olivas-gomez_commissioning_2022} and the details of the setup are shown in Fig.~2 of \cite{olivas-gomez_commissioning_2022}.
The electrically isolated target chamber acted as a Faraday cup, which was used for the integration of the beam current. To limit the buildup of carbon on the target, a copper tube cooled to liquid nitrogen temperature was placed inside the beamline, extending from the cold trap up to the target surface. The cold trap was electrically isolated and biased to -300 V in order to suppress the secondary electrons. The target was also water cooled in order to reduce the damage caused by heating due to high beam intensities.

\subsection{$\gamma$-Summing Detector: HECTOR} \label{subsec::hector}

HECTOR consists of an array of 16 $\mathrm{NaI(Tl)}$ crystals with two photo-multiplier tubes attached to each segment. Each $\mathrm{NaI(Tl)}$ crystal is housed in an aluminum casing. The crystals are arranged in a frame to form a cube 16x16x16 inches in dimension. The center of the detector array consists of a 60 mm borehole which allows the target to be placed at the center giving a 4$\pi$ total coverage for the detection of $\gamma$-rays. The advantage of using HECTOR for the $\Neag$ measurement is its high detection efficiency. The signal from each photomultiplier tube is recorded individually and then gain-matched and summed offline, to improve the detector resolution. A digital data acquisition system is used to readout the data from the 16 $\mathrm{NaI(Tl)}$ detectors. A detailed description of the HECTOR and its data-acquisition can be found in~\cite{Craig}. In this work, known level schemes were used to determine the summing efficiency of HECTOR for the measured resonances. This approach was introduced and tested for the $\Alpg$ resonances in the recent work of~\cite{olivas-gomez_commissioning_2022} and was applied in the analysis of the $\Oag$ reaction~\cite{Alex}.

\section{Experimental Procedure and Analysis} \label{sec::exp_proc_analysis}

\begin{figure}
  \centering
  \includegraphics[width=0.5\textwidth]{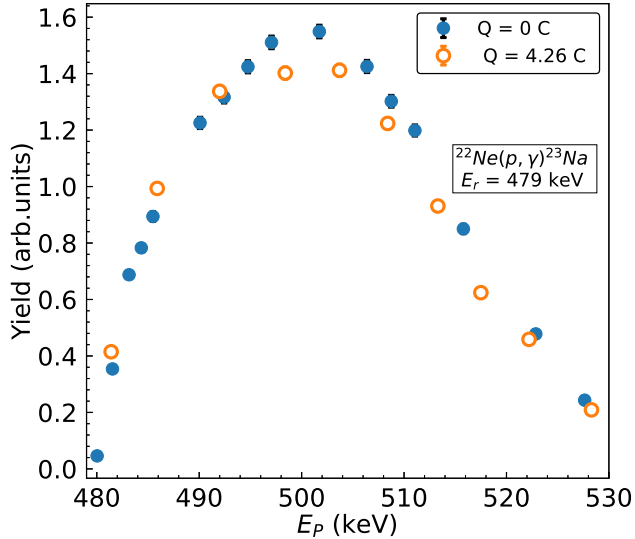}
  \caption{Yield curve for the $E_{P}^{\mathrm{lab}}$= 479 keV resonance in the $\Nepg$ reaction. The yield curve is obtained from the sum peak in HECTOR~\cite{Craig}. The filled-circle (blue) and open-circle (orange) data points show the difference in yield at the beginning and at end of the experiment. The statistical error in the data is $\simeq 1\%$, which is why the error bars are not visible. }
  \label{479_RS}
\end{figure}

The low energy resonances at $E_{\alpha}^{lab}$ = 650 and 830 keV in the $\Neag$ reaction were measured using the implanted $\Ne$ targets (as discussed in section~\ref{subsec::target}) at CASPAR~\cite{CASPAR}. The stability of the implanted $\Ne$ targets was monitored throughout the course of the experiment using the well known resonances at $E_{p}^{lab}$ = 479 keV and 851 keV in the $\Nepg$ reaction~\cite{Depalo}; the 1278 keV resonance of this reaction is not accessible with the CASPAR accelerator. Fig.~\ref{479_RS} shows the yield curve for the 479 keV resonance of $\Nepg$ for one of the $^{22}$Ne targets. The yield is obtained using the sum peak in HECTOR~\cite{Craig}, at $E_{\Sigma} = E_{CM} + Q$ = 9252.27 keV (where $E_{CM}$ is the center-of-mass energy of the projectile target system and Q is the reaction Q-value). The filled-circle (blue) and open-circle (orange) data points show the difference in the yield at the beginning and the end of the experiment, after accumulating a charge of nearly 4 C. The targets were fairly stable and the yield decreased only by 5$\%$ at the end of the experiment, as shown in Fig.~\ref{479_RS}. No heat damage or blistering was caused by the high intensity of beam current. As a consequence, we changed each target after it has been exposed to an accumulated charge of around 4 C. 

\begin{figure*}
  \centering
  \includegraphics[width=0.9\textwidth]{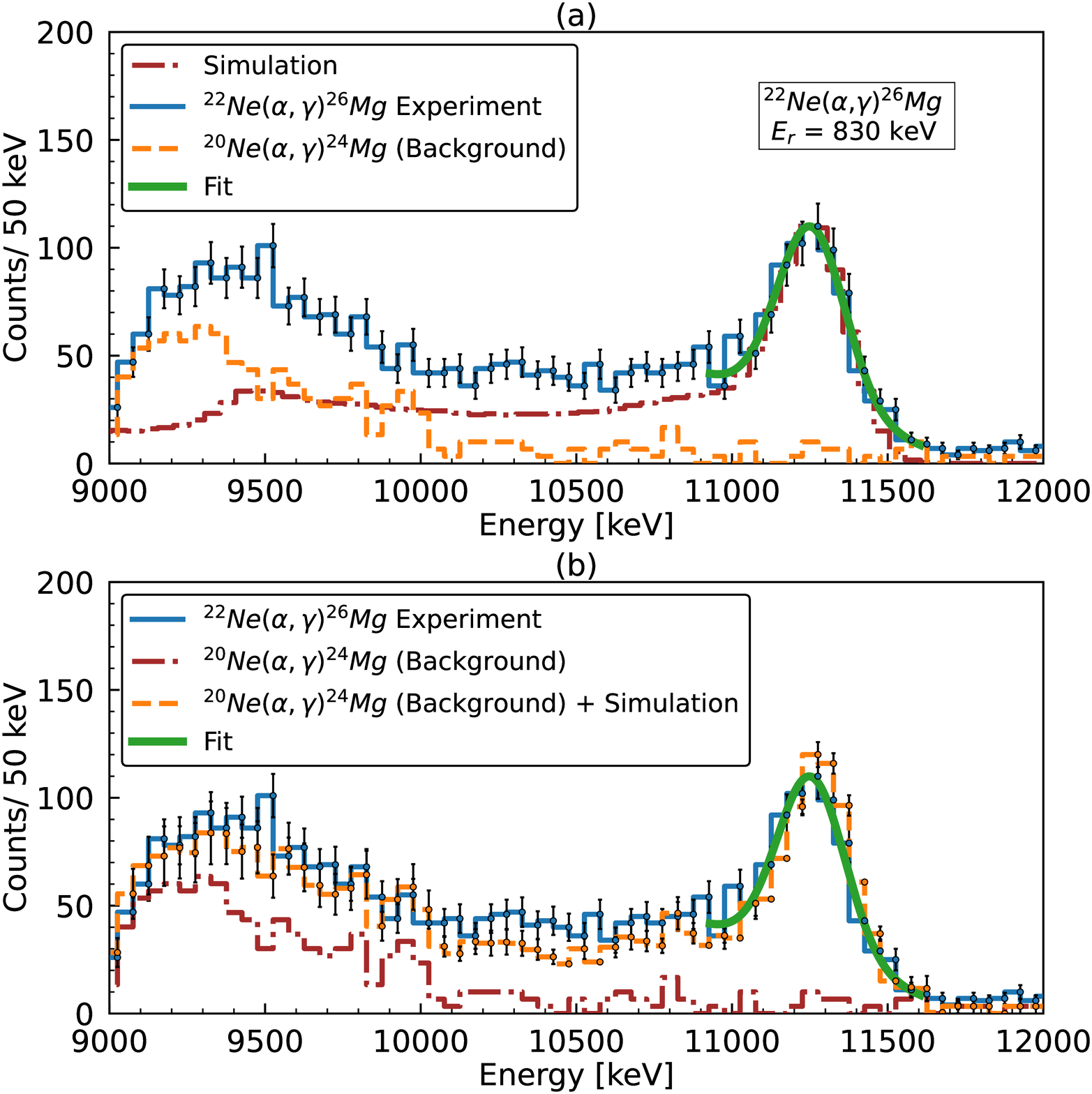}
  \caption{(a) Sum spectra for the  $E_{\alpha}^{lab}$= 830 keV resonance. The solid (blue) histogram shows the sum spectrum from the $\Neag$ reaction and the sum peak is located at $ E_{\Sigma} = E_{CM} + Q$ = 11.3 MeV. The dashed (orange) histogram shows the beam induced background using the  $^{20}$Ne target. The dash dotted (brown) histogram shows the simulated spectrum obtained from GEANT4~\cite{geant}. The thick solid (green) line shows the Gaussian plus a linear function fit to the sum peak. (b) The dashed (orange) histogram is the sum of beam induced background and simulation. It shows excellent agreement with the experimental spectrum shown in solid (blue). }
  \label{830_RS}
\end{figure*}

To optimize the experimental yield of the narrow resonance at 830 keV in $\Neag$, it has to be measured at the maximum of the $^{22}$Ne distribution. For this purpose, the energy loss of protons from the resonance energy of 479 keV till the maximum of the $\Ne$ distribution (as shown in Fig.~\ref{479_RS}) in $\Nepg$, was converted to the energy loss of $\alpha$ particles in $\Neag$ using the relative stopping powers from SRIM~\cite{SRIM}. This implies that the maximum of the $\Ne$ distribution for the $\alpha$-particle beam would correspond to an $\alpha$ energy of 910 keV for the $E_{\alpha}^{lab}$ = 830 keV resonance. The $\Ne$ concentration is constant to 10$\%$ within $\pm$ 30 keV of this energy. To confirm the position of the top of the thick target yield curve/plateau, a scan of the $E_{\alpha}^{lab}$ = 830 keV resonance was performed using large energy steps. Thus data was acquired at the top of the plateau, i.e., at $E_{\alpha}^{lab}$ = 910 keV, with a total accumulated charge of Q = 2.94 C on the $\Ne$ target. The beam-induced background was measured using an implanted $^{20}$Ne target at the same $\alpha$-particle energy. The sum peak spectrum for the 830 keV resonance is shown in Fig.~\ref{830_RS}a. The solid (blue) histogram shows the on-resonance spectrum taken with a $^{22}$Ne target, where the sum peak is located at an energy of $ E_{\Sigma}$ = 11.3 MeV. The dashed (orange) histogram shows the background spectrum obtained using a $^{20}$Ne target. The sum peak from the background $^{20}$Ne($\alpha$,$\gamma$)$^{24}$Mg reaction is located at $E_{\Sigma}$ = 10.02 MeV, which is well separated from the sum peak of $\Neag$. 

Computation of the strength of the 830 keV resonance requires the experimental yield (discussed above) as well as the efficiency of the sum peak, as evident from Equation.~\ref{eq:1}. The efficiency of the 830 keV resonance sum peak was obtained using GEANT4~\cite{geant} simulations. The level scheme and the branching ratio for the reaction were taken from the previous measurements performed by~\cite{Wolke.etal1989,Hunt.etal.19}. The entire $\gamma$-cascade was simulated and the resultant simulated spectrum is shown in Fig.~\ref{830_RS}a, as the dash dotted (brown) histogram. Adding the beam induced background to the simulation shows excellent agreement with the $\Neag$ experimental spectrum as shown in Fig.~\ref{830_RS}b.

The sum peak was fitted with a Gaussian plus a linear function shown as thick solid (green) line in Fig.~\ref{830_RS}, and was integrated within $E_{\Sigma} -3\sigma$ to $E_{\Sigma} +3\sigma$ range to get the total number of events under the peak. The simulated sum peak was fitted using the same procedure. The efficiency is given by the ratio of the number of counts in the simulated sum peak to that of the total simulated cascade events. 

The strength of the 830 keV resonance in $\Neag$  was determined relative to the well known 479 keV resonance in $\Nepg$~\cite{Depalo} using the expression for the maximum yield of an infinitely thick target~\cite{Iliadis},
\begin{equation}
         \frac{\omega \gamma_{1}}{\omega \gamma_{2}} = \left(\frac{\epsilon_{r,1}}{\epsilon_{r,2}}\right) \left(\frac{\lambda_{r,2}^{2}}{\lambda_{r,1}^{2}}\right) \left(\frac{Y_{1}}{Y_{2}}\right)\left(\frac{\eta_{2}}{\eta_{1}}\right),
         \label{rs_relative}
\end{equation}
where the subscripts 1 and 2 correspond to the resonance of interest and the standard resonance. In this approach, only relative values of the stopping powers and efficiencies are needed. The efficiency of the 479 keV resonance in $\Nepg$ ($\eta_{2}$) was obtained in the same way as 830 keV resonance, using the level scheme and branching ratio from~\cite{longland_resonance_2010}. The error in the resonance strength $\omega\gamma_{1}$ depends on the accuracy of $\omega\gamma_{2}$, and the uncertainties of the ratios of stopping powers, efficiencies and yields. Using this relative approach, we obtained $\omega\gamma$ = 35 $\pm$ 4 $\mu$eV for this resonance. The uncertainty in the resonance strength is dominated by the uncertainty of $\omega\gamma_{2}$ ($E_{P}^{lab}$ = 479 keV resonance) which is 6.4$\%$ and the relative stopping power of alphas and protons in Ta. The uncertainty in the stopping power for alphas in Ta is 6$\%$ and for protons in Ta is 3.4$\%$. The statistical uncertainty associated with the sum peak integral is 3.5$\%$ for the 830 keV resonance and 1.4$\%$ for the 479 keV resonance. The error of the relative efficiency is 5$\%$ mainly resulting from the uncertainty in the branching ratio of the 830 keV resonance in $\Neag$.

\begin{figure}
  \centering
  \includegraphics[width=0.5\textwidth]{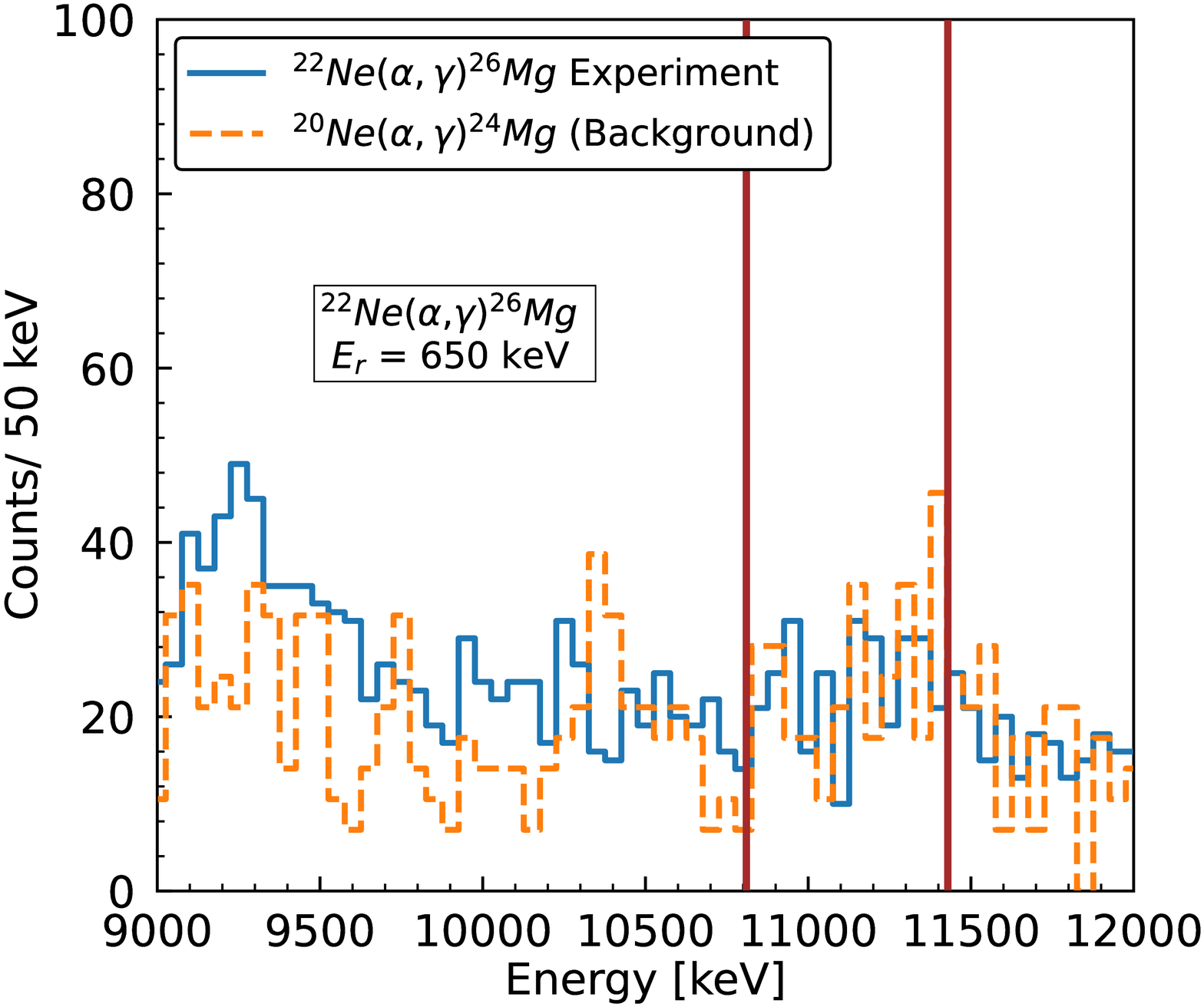}
  \caption{Sum spectra for the  $E_{\alpha}^{lab}$= 650  keV resonance. The solid (blue) histogram depicts the sum spectrum from the $\Neag$ reaction, whereas the dashed (orange) histogram shows the beam induced background using the $^{20}$Ne target. The two solid vertical lines shows the region of integration for the sum peak. The background (orange) histogram was normalized to the same accumulated charge.}
  \label{650_RS}
\end{figure}

The other low energy resonance in $\Neag$ at $E_{\alpha}^{lab}$= 650 keV was measured in the same way as the $E_{\alpha}^{lab}$= 830 keV resonance. Data was collected at $E_{\alpha}^{lab}$= 711 keV, with a total accumulated charge of Q = 22.9 C on top of the resonance. To determine the beam-induced background, data was taken with a $^{20}$Ne target at the same $\alpha$ particle energy, with a total accumulated charge of Q = 6.52 C. The sum spectrum for $E_{\alpha}^{lab}$= 650 keV resonance is shown in Fig.~\ref{650_RS}. The solid (blue) histogram shows the on resonance spectrum taken with a $\Ne$ target, where the sum peak is located at an energy of $ E_{\Sigma} $ = 11.16 MeV. The dashed (orange) histogram shows the background spectrum which was taken with the $^{20}$Ne target. The width of the region of integration was taken to be the same as the 830 keV resonance sum peak, shifting only by the center of mass energy of the incident $\alpha$-particle beam. The two vertical lines show the region of integration for the 650 keV resonance sum peak. The background on the low energy side of the peak is arising mainly from neutrons, generated by $(\alpha,n)$ reactions induced by $\alpha$-particle decay from the uranium and thorium present in the surrounding rock. The other possible source of background of high energy $\gamma$ rays is from the $^{11}$B$(\alpha,\gamma)^{15}$N reaction~\cite{11B_Wang} which has a Q-value = 10991.18 keV. $^{11}$B is usually present in the Ta backing as a contaminant. A narrow resonance in $^{11}$B$(\alpha,\gamma)^{15}$N reaction~\cite{11B_Wang} is known at an energy of 606 keV, which corresponds to an excitation energy of 11.44 MeV.

The resonance strength of the $E_{\alpha}^{lab}$= 650 keV resonance in $\Neag$ was determined relative to the  $E_{\alpha}^{lab}$= 830 keV resonance using Eq.~\ref{rs_relative}. The efficiency of the sum peak for the $E_{\alpha}^{lab}$= 650 keV resonance was first calculated taking the same decay scheme as the $E_{\alpha}^{lab}$= 830 keV resonance, since the decay scheme is unknown for this resonance. Using this relative approach, we obtained an upper limit $\omega\gamma$ $\mathrm{<0.11}$ $\mu eV$ for the $E_{\alpha}^{lab}$= 650 keV resonance. However, to calculate the influence of the choice of the involved $\gamma$-cascades on the upper limit, we simulated various cascades with different $\gamma$-ray multiplicities and $\gamma$-energies. These resulted in an upper limit of 0.15 $\mu$eV independent of the actual branching ratio of this resonance. 

During the course of the experiment, we repeatedly measured the $E_{P}^{lab}$ = 851 keV resonance in $\Nepg$. This resonance was used to monitor the target stability during the experiment, primarily because it is one of the strongest low energy resonances in $\Nepg$~\cite{endt_energy_1990}. However, the strength of the resonance is poorly known ($\pm$ 36$\%$). For this reason, we determined its resonance strength relative to the known $E_{P}^{lab}$ = 479 keV resonance ($\omega\gamma$ = 0.594 $\pm$ 0.038 eV)~\cite{Depalo} using Eq.~\ref{rs_relative}. The level scheme and branching ratio to calculate the efficiency of the 851 keV resonance sum peak were taken from~\cite{BAKKUM19891}. We obtained a resonance strength of $\omega\gamma$ = 9.2 ± 0.7 eV for this resonance. This value is in agreement with the previously reported value of $\omega\gamma$ = 7.0 ± 2.5 eV~\cite{endt_energy_1990} but with significantly lower uncertainty. With this improved accuracy this strong low energy resonance can now be used to trace Ne implanted in various materials with high sensitivity without the need for proton energies past 1 MeV.

\begin{table}

\caption{Comparison of the previous literature resonance strength values with the present work for the $E_{\alpha}^{lab}$= 830 keV resonance\label{table1}}
\begin{ruledtabular}
\begin{tabular}{c c}
Work & $\mathrm{\omega\gamma[\mu eV]}$ \\
\hline
Wolke \textit{et al.} \cite{Wolke.etal1989} & 36 $\pm$ 4 \\
Jaeger (Thesis) \cite{Jaeger} & 33 $\pm$ 4\\
Hunt \textit{et al.} \cite{Hunt.etal.19} & 46 $\pm$ 12 \\
This work & 35 $\pm$ 4 \\
\hline
Weighted average\footnote{Common uncertainties of the individual values are negligible (for details see text)}   & 35 $\pm$ 2  \\


\end{tabular}
 \end{ruledtabular}
 
\end{table}

\section{Discussion}

The present resonance strength for the 830 keV resonance is compared with the results of previous measurements in Table.~\ref{table1}. Wolke \textit{et al.}~\cite{Wolke.etal1989} used one passively shielded Ge detector whereas Jaeger \textit{et al.}~\cite{Jaeger} used one passively and actively shielded Ge detector. Both of these measurements used windowless gas cell target with enriched $\Ne$ gas. Hunt \textit{et al.}~\cite{Hunt.etal.19} also used one passively and actively shielded detector and the target was produced by implantation of $\Ne$ onto Titanium. In contrast to the previous measurements, our present experiment utilized a $\gamma$-summing detector which was located on the 4850 ft. level of the Sanford Underground Research Facility in Lead, South Dakota~\cite{CASPAR} which reduced cosmic ray induced background significantly. Here, the targets consisted of $\Ne$ implanted onto Ta sheets. A common uncertainty of these experiments is the $\alpha$ stopping power obtained from SRIM~\cite{SRIM}. The uncertainty in the alpha stopping power for pure $\Ne$ gas is ~2$\%$ which is small compared to the total error of ~10$\%$ of the two gas target experiments \cite{Wolke.etal1989, Jaeger}. Titanium backings were used by Hunt \textit{et al.} which accounted for an uncertainty of 5$\%$ for the alpha stopping power which is small compared to the total error of 26$\%$. The present experiment uses Ta backings which have an uncertainty of 5$\%$ for the alpha stopping power. Our present measurement is relative to the $\Nepg$ 479 keV resonance and we never make use of the absolute value of the stopping power. Only the ratio of alpha to proton stopping power is required. For these reasons common uncertainties of the individual experimental values are negligible. This leads to a weighted average of the previous and present measurements of $\omega\gamma$ = 35 $\pm$ 2 $\mu eV$.

While we confirm the previous studies of the resonance strength for the 830 keV resonance in $\Neag$, the strength of the corresponding resonance in the $\Nean$ still carries large uncertainties. Previous direct studies of this resonance in $\Nean$~\cite{Jaeger, Harms, Giesen} suggests values for its resonance strength which differ substantially (up to a factor of four) amongst each other. These measurements also differ from the prediction of the coincidence studies of the decay of this level using the $^{22}$Ne$(^6$Li$,d\gamma)$ and the $^{22}$Ne$(^6$Li$,d n)$ reactions~\cite{ota_decay_2020}. Unlike the direct measurements the coincidence studies suggest a comparable strength between the $(\alpha,n)$ and $(\alpha,\gamma)$ decay modes. Further studies of the $(\alpha,n)$ reactions are needed to constrain the resonance strengths.

For the lowest energy resonance at $E_{\alpha}^{lab}$ = 650 keV, the observed yield only provides an upper limit for the resonance strength ($\omega\gamma$ $\mathrm {<0.15}$ $\mu eV$). This limit is a factor of ~4 lower than suggested by~\cite{Talwar.etal2016} on the basis of a J$^{\pi}=1^{-}$ assignment obtained from the angular distribution analysis of the $^{22}$Ne($^6$Li,d)$^{26}$Mg and $^{26}$Mg($\alpha,\alpha')^{26}$Mg studies. Low energy studies of the $^{22}$Ne($^6$Li,d)$^{26}$Mg and $^{22}$Ne($^7$Li,t)$^{26}$Mg reactions~\cite{JAYATISSA} on the other hand observed the $E_{x}$ = 11.3 MeV ($E_{\alpha}^{lab}$ =  830 keV) state but did not populate the state at $E_{x}$ = 11.17 MeV ($E_{\alpha}^{lab}$ =  650 keV). The measurement performed by~\cite{Talwar.etal2016} was at about twice the center of mass energy of~\cite{ota_decay_2020}, whereas~\cite{JAYATISSA} did the measurement at sub-Coulomb energies. For this reason it could be possible that~\cite{Talwar.etal2016} observed a high spin state while this is unlikely for the low energy experiments performed by~\cite{ota_decay_2020} and \cite{JAYATISSA}. This might suggest that the state at $E_{x}$ = 11.17 MeV ($E_{\alpha}^{lab}$ = 650 keV), is a high spin state, e.g. J$^{\pi}=4^{+}$, which cannot be excluded from angular distributions~\cite{Talwar.etal2016}. The single particle width ( $\Gamma_{sp} = 5 \times 10^{-10} $ eV) is three orders of magnitude lower than that of a J$^{\pi}=1^{-}$ state ( $\Gamma_{sp} = 5 \times 10^{-7} $ eV), predicted by~\cite{Talwar.etal2016}. Several states have been observed in this excitation range of $^{26}$Mg, which cannot be resolved in the transfer and scattering studies discussed here, so it cannot be excluded that the $\alpha$ elastic scattering populates a different state than the $\alpha$ transfer reactions. The $\alpha$ capture data discussed here indicate that the state at $E_{x}$ = 11.17 MeV is a high spin state and for this reason should have a distinctively smaller resonance strength than predicted by~\cite{Talwar.etal2016} based on a 1$^-$ assignment. 

If the resonance at $E_{\alpha}^{lab}$ = 650 keV would indeed correspond to a high spin state, 4$^+$, then its contribution to the $\Neag$ reaction rate would be reduced by two orders of magnitude because of the reduction of the penetrability. In this case its contribution to the total ($\alpha,\gamma$) reaction rate would be less significant and would only play an important role in the temperature range around 0.2 GK (see e.g. Figure 8 of \cite{Talwar.etal2016}). At higher temperatures the reaction rate would be mostly dominated by the resonance at $E_{\alpha}^{lab}$ = 830 keV and towards lower temperatures by resonances corresponding to neutron bound states in $^{26}$Mg. As a consequence the depletion of $^{22}$Ne during the lower temperature burning phases between helium flashes in the AGB phase and the cooler phases of core helium burning would be reduced. This in turn would translate into an enhanced production of s-process elements which would be increased compared to the network calculation of \cite{Talwar.etal2016} (see e.g. Figures 14 and 17 of \cite{Talwar.etal2016}).

 \section{Summary}
 In summary, we measured the two low energy resonances in $\Neag$, which dominate the reaction rate at stellar helium burning temperatures around 0.3 GK. The measurement of the $E_{\alpha}^{lab}$ = 830 keV resonance confirms the previous results with considerably smaller uncertainties. Based on our present measurement we provide a  new weighted average value of the resonance strength, which is 35 $\pm$ 2 $\mu eV$. We performed the first direct measurement of the lowest energy resonance at $E_{\alpha}^{lab}$ = 650 keV. We provide an upper limit for the resonance strength which is considerably lower than suggested by~\cite{Talwar.etal2016}. For future measurements, either an extended gas target could be used in an underground environment, which would improve the yield substantially compared to using the implanted $\Ne$ target or an experiment could be performed using inverse kinematics and a recoil separator. However, considering the discussion of the properties of this resonance a direct measurement might not be able to observe this resonance if it is indeed a high spin state. In addition, we also measured the $E_{P}^{lab}$= 851 keV resonance in $\Nepg$, and obtained a resonance strength $\omega\gamma$ = 9.2 $\pm$ 0.7 eV, with significantly lower uncertainties compared to previous measurements.

\begin{acknowledgments}
The authors would like to thank the staff of the Sanford Underground Research Facility (SURF) for their hospitality and technical support given during the course of the experiment.
 This research was funded by the National Science Foundation through Grant No.~PHY-2011890 (University of Notre Dame Nuclear Science Laboratory), Grant No.~PHY-1430152 (the Joint Institute for Nuclear Astrophysics - Center for the Evolution of the Elements), Grant No.~PHY-1913746 (The Compact Accelerator System for Performing Astrophysical Research at the Sanford Underground Research Facility), the Sanford Underground Research Facility (SURF, Grant No.~DE-SC0020216) and Grant No.~PHY-1614442. 
\end{acknowledgments}

\bibliography{22Ne.bib}

\begin{thebibliography}{30}%
\makeatletter
\providecommand \@ifxundefined [1]{%
 \@ifx{#1\undefined}
}%
\providecommand \@ifnum [1]{%
 \ifnum #1\expandafter \@firstoftwo
 \else \expandafter \@secondoftwo
 \fi
}%
\providecommand \@ifx [1]{%
 \ifx #1\expandafter \@firstoftwo
 \else \expandafter \@secondoftwo
 \fi
}%
\providecommand \natexlab [1]{#1}%
\providecommand \enquote  [1]{``#1''}%
\providecommand \bibnamefont  [1]{#1}%
\providecommand \bibfnamefont [1]{#1}%
\providecommand \citenamefont [1]{#1}%
\providecommand \href@noop [0]{\@secondoftwo}%
\providecommand \href [0]{\begingroup \@sanitize@url \@href}%
\providecommand \@href[1]{\@@startlink{#1}\@@href}%
\providecommand \@@href[1]{\endgroup#1\@@endlink}%
\providecommand \@sanitize@url [0]{\catcode `\\12\catcode `\$12\catcode
  `\&12\catcode `\#12\catcode `\^12\catcode `\_12\catcode `\%12\relax}%
\providecommand \@@startlink[1]{}%
\providecommand \@@endlink[0]{}%
\providecommand \url  [0]{\begingroup\@sanitize@url \@url }%
\providecommand \@url [1]{\endgroup\@href {#1}{\urlprefix }}%
\providecommand \urlprefix  [0]{URL }%
\providecommand \Eprint [0]{\href }%
\providecommand \doibase [0]{http://dx.doi.org/}%
\providecommand \selectlanguage [0]{\@gobble}%
\providecommand \bibinfo  [0]{\@secondoftwo}%
\providecommand \bibfield  [0]{\@secondoftwo}%
\providecommand \translation [1]{[#1]}%
\providecommand \BibitemOpen [0]{}%
\providecommand \bibitemStop [0]{}%
\providecommand \bibitemNoStop [0]{.\EOS\space}%
\providecommand \EOS [0]{\spacefactor3000\relax}%
\providecommand \BibitemShut  [1]{\csname bibitem#1\endcsname}%
\let\auto@bib@innerbib\@empty
\bibitem [{\citenamefont {{Rolfs}}\ \emph {et~al.}(1989)\citenamefont
  {{Rolfs}}, \citenamefont {{Rodney}},\ and\ \citenamefont
  {{Clayton}}}]{Rolfs.etal.89}%
  \BibitemOpen
  \bibfield  {author} {\bibinfo {author} {\bibfnamefont {C.~E.}\ \bibnamefont
  {{Rolfs}}}, \bibinfo {author} {\bibfnamefont {W.~S.}\ \bibnamefont
  {{Rodney}}}, \ and\ \bibinfo {author} {\bibfnamefont {D.~D.}\ \bibnamefont
  {{Clayton}}},\ }\href {\doibase 10.1119/1.16074} {\bibfield  {journal}
  {\bibinfo  {journal} {American Journal of Physics}\ }\textbf {\bibinfo
  {volume} {57}},\ \bibinfo {pages} {188} (\bibinfo {year} {1989})}\BibitemShut
  {NoStop}%
\bibitem [{\citenamefont {Bisterzo}\ \emph {et~al.}(2015)\citenamefont
  {Bisterzo}, \citenamefont {Gallino}, \citenamefont {Käppeler}, \citenamefont
  {Wiescher}, \citenamefont {Imbriani}, \citenamefont {Straniero},
  \citenamefont {Cristallo}, \citenamefont {Görres},\ and\ \citenamefont
  {deBoer}}]{bisterzo_branchings_2015}%
  \BibitemOpen
  \bibfield  {author} {\bibinfo {author} {\bibfnamefont {S.}~\bibnamefont
  {Bisterzo}}, \bibinfo {author} {\bibfnamefont {R.}~\bibnamefont {Gallino}},
  \bibinfo {author} {\bibfnamefont {F.}~\bibnamefont {Käppeler}}, \bibinfo
  {author} {\bibfnamefont {M.}~\bibnamefont {Wiescher}}, \bibinfo {author}
  {\bibfnamefont {G.}~\bibnamefont {Imbriani}}, \bibinfo {author}
  {\bibfnamefont {O.}~\bibnamefont {Straniero}}, \bibinfo {author}
  {\bibfnamefont {S.}~\bibnamefont {Cristallo}}, \bibinfo {author}
  {\bibfnamefont {J.}~\bibnamefont {Görres}}, \ and\ \bibinfo {author}
  {\bibfnamefont {R.~J.}\ \bibnamefont {deBoer}},\ }\href {\doibase
  10.1093/mnras/stv271} {\bibfield  {journal} {\bibinfo  {journal} {Monthly
  Notices of the Royal Astronomical Society}\ }\textbf {\bibinfo {volume}
  {449}},\ \bibinfo {pages} {506} (\bibinfo {year} {2015})},\ \bibinfo {note}
  {\_eprint:
  https://academic.oup.com/mnras/article-pdf/449/1/506/13767772/stv271.pdf}\BibitemShut
  {NoStop}%
\bibitem [{\citenamefont {{K\"appeler}}\ \emph {et~al.}(1994)\citenamefont
  {{K\"appeler}}, \citenamefont {{Wiescher}}, \citenamefont {{Giesen}},
  \citenamefont {{G\"orres}}, \citenamefont {{Baraffe}}, \citenamefont {{El
  Eid}}, \citenamefont {{Raiteri}}, \citenamefont {{Busso}}, \citenamefont
  {{Gallino}}, \citenamefont {{Limongi}},\ and\ \citenamefont
  {{Chieffi}}}]{Kaeppeler}%
  \BibitemOpen
  \bibfield  {author} {\bibinfo {author} {\bibfnamefont {F.}~\bibnamefont
  {{K\"appeler}}}, \bibinfo {author} {\bibfnamefont {M.}~\bibnamefont
  {{Wiescher}}}, \bibinfo {author} {\bibfnamefont {U.}~\bibnamefont
  {{Giesen}}}, \bibinfo {author} {\bibfnamefont {J.}~\bibnamefont
  {{G\"orres}}}, \bibinfo {author} {\bibfnamefont {I.}~\bibnamefont
  {{Baraffe}}}, \bibinfo {author} {\bibfnamefont {M.}~\bibnamefont {{El Eid}}},
  \bibinfo {author} {\bibfnamefont {C.~M.}\ \bibnamefont {{Raiteri}}}, \bibinfo
  {author} {\bibfnamefont {M.}~\bibnamefont {{Busso}}}, \bibinfo {author}
  {\bibfnamefont {R.}~\bibnamefont {{Gallino}}}, \bibinfo {author}
  {\bibfnamefont {M.}~\bibnamefont {{Limongi}}}, \ and\ \bibinfo {author}
  {\bibfnamefont {A.}~\bibnamefont {{Chieffi}}},\ }\href {\doibase
  10.1086/175004} {\bibfield  {journal} {\bibinfo  {journal} {\apj}\ }\textbf
  {\bibinfo {volume} {437}},\ \bibinfo {pages} {396} (\bibinfo {year}
  {1994})}\BibitemShut {NoStop}%
\bibitem [{\citenamefont {Talwar}\ \emph {et~al.}(2016)\citenamefont {Talwar},
  \citenamefont {Adachi}, \citenamefont {Berg}, \citenamefont {Bin},
  \citenamefont {Bisterzo}, \citenamefont {Couder}, \citenamefont {deBoer},
  \citenamefont {Fang}, \citenamefont {Fujita}, \citenamefont {Fujita},
  \citenamefont {G\"orres}, \citenamefont {Hatanaka}, \citenamefont {Itoh},
  \citenamefont {Kadoya}, \citenamefont {Long}, \citenamefont {Miki},
  \citenamefont {Patel}, \citenamefont {Pignatari}, \citenamefont {Shimbara},
  \citenamefont {Tamii}, \citenamefont {Wiescher}, \citenamefont {Yamamoto},\
  and\ \citenamefont {Yosoi}}]{Talwar.etal2016}%
  \BibitemOpen
  \bibfield  {author} {\bibinfo {author} {\bibfnamefont {R.}~\bibnamefont
  {Talwar}}, \bibinfo {author} {\bibfnamefont {T.}~\bibnamefont {Adachi}},
  \bibinfo {author} {\bibfnamefont {G.~P.~A.}\ \bibnamefont {Berg}}, \bibinfo
  {author} {\bibfnamefont {L.}~\bibnamefont {Bin}}, \bibinfo {author}
  {\bibfnamefont {S.}~\bibnamefont {Bisterzo}}, \bibinfo {author}
  {\bibfnamefont {M.}~\bibnamefont {Couder}}, \bibinfo {author} {\bibfnamefont
  {R.~J.}\ \bibnamefont {deBoer}}, \bibinfo {author} {\bibfnamefont
  {X.}~\bibnamefont {Fang}}, \bibinfo {author} {\bibfnamefont {H.}~\bibnamefont
  {Fujita}}, \bibinfo {author} {\bibfnamefont {Y.}~\bibnamefont {Fujita}},
  \bibinfo {author} {\bibfnamefont {J.}~\bibnamefont {G\"orres}}, \bibinfo
  {author} {\bibfnamefont {K.}~\bibnamefont {Hatanaka}}, \bibinfo {author}
  {\bibfnamefont {T.}~\bibnamefont {Itoh}}, \bibinfo {author} {\bibfnamefont
  {T.}~\bibnamefont {Kadoya}}, \bibinfo {author} {\bibfnamefont
  {A.}~\bibnamefont {Long}}, \bibinfo {author} {\bibfnamefont {K.}~\bibnamefont
  {Miki}}, \bibinfo {author} {\bibfnamefont {D.}~\bibnamefont {Patel}},
  \bibinfo {author} {\bibfnamefont {M.}~\bibnamefont {Pignatari}}, \bibinfo
  {author} {\bibfnamefont {Y.}~\bibnamefont {Shimbara}}, \bibinfo {author}
  {\bibfnamefont {A.}~\bibnamefont {Tamii}}, \bibinfo {author} {\bibfnamefont
  {M.}~\bibnamefont {Wiescher}}, \bibinfo {author} {\bibfnamefont
  {T.}~\bibnamefont {Yamamoto}}, \ and\ \bibinfo {author} {\bibfnamefont
  {M.}~\bibnamefont {Yosoi}},\ }\href {\doibase 10.1103/PhysRevC.93.055803}
  {\bibfield  {journal} {\bibinfo  {journal} {Phys. Rev. C}\ }\textbf {\bibinfo
  {volume} {93}},\ \bibinfo {pages} {055803} (\bibinfo {year}
  {2016})}\BibitemShut {NoStop}%
\bibitem [{\citenamefont {{Wolke}}\ \emph {et~al.}(1989)\citenamefont
  {{Wolke}}, \citenamefont {{Becker}}, \citenamefont {{Rolfs}}, \citenamefont
  {{Schr{\"o}der}}, \citenamefont {{Trautvetter}}, \citenamefont {{Harms}},
  \citenamefont {{Kratz}}, \citenamefont {{Hammer}}, \citenamefont
  {{Wiescher}},\ and\ \citenamefont {{W{\"o}hr}}}]{Wolke.etal1989}%
  \BibitemOpen
  \bibfield  {author} {\bibinfo {author} {\bibfnamefont {K.}~\bibnamefont
  {{Wolke}}}, \bibinfo {author} {\bibfnamefont {H.~W.}\ \bibnamefont
  {{Becker}}}, \bibinfo {author} {\bibfnamefont {C.}~\bibnamefont {{Rolfs}}},
  \bibinfo {author} {\bibfnamefont {U.}~\bibnamefont {{Schr{\"o}der}}},
  \bibinfo {author} {\bibfnamefont {H.~P.}\ \bibnamefont {{Trautvetter}}},
  \bibinfo {author} {\bibfnamefont {V.}~\bibnamefont {{Harms}}}, \bibinfo
  {author} {\bibfnamefont {K.~L.}\ \bibnamefont {{Kratz}}}, \bibinfo {author}
  {\bibfnamefont {J.~W.}\ \bibnamefont {{Hammer}}}, \bibinfo {author}
  {\bibfnamefont {M.}~\bibnamefont {{Wiescher}}}, \ and\ \bibinfo {author}
  {\bibfnamefont {A.}~\bibnamefont {{W{\"o}hr}}},\ }\href {\doibase
  10.1007/BF01294757} {\bibfield  {journal} {\bibinfo  {journal} {Zeitschrift
  fur Physik A Hadrons and Nuclei}\ }\textbf {\bibinfo {volume} {334}},\
  \bibinfo {pages} {491} (\bibinfo {year} {1989})}\BibitemShut {NoStop}%
\bibitem [{\citenamefont {{Hunt}}\ \emph {et~al.}(2019)\citenamefont {{Hunt}},
  \citenamefont {{Iliadis}}, \citenamefont {{Champagne}}, \citenamefont
  {{Downen}},\ and\ \citenamefont {{Cooper}}}]{Hunt.etal.19}%
  \BibitemOpen
  \bibfield  {author} {\bibinfo {author} {\bibfnamefont {S.}~\bibnamefont
  {{Hunt}}}, \bibinfo {author} {\bibfnamefont {C.}~\bibnamefont {{Iliadis}}},
  \bibinfo {author} {\bibfnamefont {A.}~\bibnamefont {{Champagne}}}, \bibinfo
  {author} {\bibfnamefont {L.}~\bibnamefont {{Downen}}}, \ and\ \bibinfo
  {author} {\bibfnamefont {A.}~\bibnamefont {{Cooper}}},\ }\href {\doibase
  10.1103/PhysRevC.99.045804} {\bibfield  {journal} {\bibinfo  {journal}
  {\prc}\ }\textbf {\bibinfo {volume} {99}},\ \bibinfo {eid} {045804} (\bibinfo
  {year} {2019})}\BibitemShut {NoStop}%
\bibitem [{\citenamefont {Selin}\ \emph {et~al.}(1967)\citenamefont {Selin},
  \citenamefont {Arnell},\ and\ \citenamefont {Almén}}]{SELIN1967218}%
  \BibitemOpen
  \bibfield  {author} {\bibinfo {author} {\bibfnamefont {E.}~\bibnamefont
  {Selin}}, \bibinfo {author} {\bibfnamefont {S.}~\bibnamefont {Arnell}}, \
  and\ \bibinfo {author} {\bibfnamefont {O.}~\bibnamefont {Almén}},\ }\href
  {\doibase https://doi.org/10.1016/0029-554X(67)90193-0} {\bibfield  {journal}
  {\bibinfo  {journal} {Nuclear Instruments and Methods}\ }\textbf {\bibinfo
  {volume} {56}},\ \bibinfo {pages} {218} (\bibinfo {year} {1967})}\BibitemShut
  {NoStop}%
\bibitem [{\citenamefont {{Depalo}}\ \emph {et~al.}(2015)\citenamefont
  {{Depalo}}, \citenamefont {{Cavanna}}, \citenamefont {{Ferraro}},
  \citenamefont {{Slemer}}, \citenamefont {{Al-Abdullah}}, \citenamefont
  {{Akhmadaliev}}, \citenamefont {{Anders}}, \citenamefont {{Bemmerer}},
  \citenamefont {{Elekes}}, \citenamefont {{Mattei}}, \citenamefont
  {{Reinicke}}, \citenamefont {{Schmidt}}, \citenamefont {{Scian}},\ and\
  \citenamefont {{Wagner}}}]{Depalo}%
  \BibitemOpen
  \bibfield  {author} {\bibinfo {author} {\bibfnamefont {R.}~\bibnamefont
  {{Depalo}}}, \bibinfo {author} {\bibfnamefont {F.}~\bibnamefont {{Cavanna}}},
  \bibinfo {author} {\bibfnamefont {F.}~\bibnamefont {{Ferraro}}}, \bibinfo
  {author} {\bibfnamefont {A.}~\bibnamefont {{Slemer}}}, \bibinfo {author}
  {\bibfnamefont {T.}~\bibnamefont {{Al-Abdullah}}}, \bibinfo {author}
  {\bibfnamefont {S.}~\bibnamefont {{Akhmadaliev}}}, \bibinfo {author}
  {\bibfnamefont {M.}~\bibnamefont {{Anders}}}, \bibinfo {author}
  {\bibfnamefont {D.}~\bibnamefont {{Bemmerer}}}, \bibinfo {author}
  {\bibfnamefont {Z.}~\bibnamefont {{Elekes}}}, \bibinfo {author}
  {\bibfnamefont {G.}~\bibnamefont {{Mattei}}}, \bibinfo {author}
  {\bibfnamefont {S.}~\bibnamefont {{Reinicke}}}, \bibinfo {author}
  {\bibfnamefont {K.}~\bibnamefont {{Schmidt}}}, \bibinfo {author}
  {\bibfnamefont {C.}~\bibnamefont {{Scian}}}, \ and\ \bibinfo {author}
  {\bibfnamefont {L.}~\bibnamefont {{Wagner}}},\ }\href {\doibase
  10.1103/PhysRevC.92.045807} {\bibfield  {journal} {\bibinfo  {journal}
  {\prc}\ }\textbf {\bibinfo {volume} {92}},\ \bibinfo {eid} {045807} (\bibinfo
  {year} {2015})},\ \Eprint {http://arxiv.org/abs/1507.03893} {arXiv:1507.03893
  [nucl-ex]} \BibitemShut {NoStop}%
\bibitem [{\citenamefont {{Antilla}}\ \emph {et~al.}(1977)\citenamefont
  {{Antilla}}, \citenamefont {{Keinonen}}, \citenamefont {{Hautala}},\ and\
  \citenamefont {{Forsblom}}}]{Antilla}%
  \BibitemOpen
  \bibfield  {author} {\bibinfo {author} {\bibfnamefont {A.}~\bibnamefont
  {{Antilla}}}, \bibinfo {author} {\bibfnamefont {J.}~\bibnamefont
  {{Keinonen}}}, \bibinfo {author} {\bibfnamefont {M.}~\bibnamefont
  {{Hautala}}}, \ and\ \bibinfo {author} {\bibfnamefont {I.}~\bibnamefont
  {{Forsblom}}},\ }\href {\doibase 10.1016/0029-554X(77)90393-7} {\bibfield
  {journal} {\bibinfo  {journal} {Nuclear Instruments and Methods}\ }\textbf
  {\bibinfo {volume} {147}},\ \bibinfo {pages} {501} (\bibinfo {year}
  {1977})}\BibitemShut {NoStop}%
\bibitem [{\citenamefont {Paine}\ and\ \citenamefont
  {Sargood}(1979)}]{PAINE1979389}%
  \BibitemOpen
  \bibfield  {author} {\bibinfo {author} {\bibfnamefont {B.}~\bibnamefont
  {Paine}}\ and\ \bibinfo {author} {\bibfnamefont {D.}~\bibnamefont
  {Sargood}},\ }\href {\doibase https://doi.org/10.1016/0375-9474(79)90349-X}
  {\bibfield  {journal} {\bibinfo  {journal} {Nuclear Physics A}\ }\textbf
  {\bibinfo {volume} {331}},\ \bibinfo {pages} {389} (\bibinfo {year}
  {1979})}\BibitemShut {NoStop}%
\bibitem [{\citenamefont {{Iliadis}}(2007)}]{Iliadis}%
  \BibitemOpen
  \bibfield  {author} {\bibinfo {author} {\bibfnamefont {C.}~\bibnamefont
  {{Iliadis}}},\ }\href {\doibase 10.1002/9783527692668} {\emph {\bibinfo
  {title} {{Nuclear Physics of Stars}}}}\ (\bibinfo {year} {2007})\BibitemShut
  {NoStop}%
\bibitem [{\citenamefont {Ziegler}\ \emph {et~al.}(2010)\citenamefont
  {Ziegler}, \citenamefont {Ziegler},\ and\ \citenamefont {Biersack}}]{SRIM}%
  \BibitemOpen
  \bibfield  {author} {\bibinfo {author} {\bibfnamefont {J.~F.}\ \bibnamefont
  {Ziegler}}, \bibinfo {author} {\bibfnamefont {M.}~\bibnamefont {Ziegler}}, \
  and\ \bibinfo {author} {\bibfnamefont {J.}~\bibnamefont {Biersack}},\ }\href
  {\doibase https://doi.org/10.1016/j.nimb.2010.02.091} {\bibfield  {journal}
  {\bibinfo  {journal} {Nuclear Instruments and Methods in Physics Research
  Section B: Beam Interactions with Materials and Atoms}\ }\textbf {\bibinfo
  {volume} {268}},\ \bibinfo {pages} {1818} (\bibinfo {year} {2010})},\
  \bibinfo {note} {19th International Conference on Ion Beam
  Analysis}\BibitemShut {NoStop}%
\bibitem [{\citenamefont {Bakkum}\ and\ \citenamefont {{Van Der
  Leun}}(1989)}]{BAKKUM19891}%
  \BibitemOpen
  \bibfield  {author} {\bibinfo {author} {\bibfnamefont {E.}~\bibnamefont
  {Bakkum}}\ and\ \bibinfo {author} {\bibfnamefont {C.}~\bibnamefont {{Van Der
  Leun}}},\ }\href {\doibase https://doi.org/10.1016/0375-9474(89)90128-0}
  {\bibfield  {journal} {\bibinfo  {journal} {Nuclear Physics A}\ }\textbf
  {\bibinfo {volume} {500}},\ \bibinfo {pages} {1} (\bibinfo {year}
  {1989})}\BibitemShut {NoStop}%
\bibitem [{\citenamefont {Viitasalo}\ \emph {et~al.}(1972)\citenamefont
  {Viitasalo}, \citenamefont {Piiparinen},\ and\ \citenamefont
  {Anttila}}]{viitasalo_angular_1972}%
  \BibitemOpen
  \bibfield  {author} {\bibinfo {author} {\bibfnamefont {M.}~\bibnamefont
  {Viitasalo}}, \bibinfo {author} {\bibfnamefont {M.}~\bibnamefont
  {Piiparinen}}, \ and\ \bibinfo {author} {\bibfnamefont {A.}~\bibnamefont
  {Anttila}},\ }\href {\doibase 10.1007/BF01379751} {\bibfield  {journal}
  {\bibinfo  {journal} {Zeitschrift für Physik A Hadrons and nuclei}\ }\textbf
  {\bibinfo {volume} {250}},\ \bibinfo {pages} {387} (\bibinfo {year}
  {1972})}\BibitemShut {NoStop}%
\bibitem [{\citenamefont {{Robertson}}\ \emph {et~al.}(2016)\citenamefont
  {{Robertson}}, \citenamefont {{Couder}}, \citenamefont {{Greife}},
  \citenamefont {{Strieder}},\ and\ \citenamefont {{Wiescher}}}]{CASPAR}%
  \BibitemOpen
  \bibfield  {author} {\bibinfo {author} {\bibfnamefont {D.}~\bibnamefont
  {{Robertson}}}, \bibinfo {author} {\bibfnamefont {M.}~\bibnamefont
  {{Couder}}}, \bibinfo {author} {\bibfnamefont {U.}~\bibnamefont {{Greife}}},
  \bibinfo {author} {\bibfnamefont {F.}~\bibnamefont {{Strieder}}}, \ and\
  \bibinfo {author} {\bibfnamefont {M.}~\bibnamefont {{Wiescher}}},\ }in\ \href
  {\doibase 10.1051/epjconf/201610909002} {\emph {\bibinfo {booktitle}
  {European Physical Journal Web of Conferences}}},\ \bibinfo {series}
  {European Physical Journal Web of Conferences}, Vol.\ \bibinfo {volume}
  {109}\ (\bibinfo {year} {2016})\ p.\ \bibinfo {pages} {09002}\BibitemShut
  {NoStop}%
\bibitem [{cas()}]{caspar_site}%
  \BibitemOpen
  \href@noop {} {\enquote {\bibinfo {title} {Compact accelerator system for
  performing astrophysical research},}\ }\bibinfo {howpublished}
  {\url{https://caspar.nd.edu/}},\ \bibinfo {note} {accessed:
  2022-04-01}\BibitemShut {NoStop}%
\bibitem [{\citenamefont {Vogelaar}\ \emph {et~al.}(1990)\citenamefont
  {Vogelaar}, \citenamefont {Wang}, \citenamefont {Kellogg},\ and\
  \citenamefont {Kavanagh}}]{Vogelaar}%
  \BibitemOpen
  \bibfield  {author} {\bibinfo {author} {\bibfnamefont {R.~B.}\ \bibnamefont
  {Vogelaar}}, \bibinfo {author} {\bibfnamefont {T.~R.}\ \bibnamefont {Wang}},
  \bibinfo {author} {\bibfnamefont {S.~E.}\ \bibnamefont {Kellogg}}, \ and\
  \bibinfo {author} {\bibfnamefont {R.~W.}\ \bibnamefont {Kavanagh}},\ }\href
  {\doibase 10.1103/PhysRevC.42.753} {\bibfield  {journal} {\bibinfo  {journal}
  {Phys. Rev. C}\ }\textbf {\bibinfo {volume} {42}},\ \bibinfo {pages} {753}
  (\bibinfo {year} {1990})}\BibitemShut {NoStop}%
\bibitem [{\citenamefont {Endt}\ and\ \citenamefont {van~dèr
  Leun}(1973)}]{osti_4398955}%
  \BibitemOpen
  \bibfield  {author} {\bibinfo {author} {\bibfnamefont {P.~M.}\ \bibnamefont
  {Endt}}\ and\ \bibinfo {author} {\bibfnamefont {C.}~\bibnamefont {van~dèr
  Leun}},\ }\href {\doibase 10.1016/0375-9474(73)91131-7} {\bibfield  {journal}
  {\bibinfo  {journal} {Nuclear Physics. A}\ }\textbf {\bibinfo {volume} {214}}
  (\bibinfo {year} {1973}),\ 10.1016/0375-9474(73)91131-7}\BibitemShut
  {NoStop}%
\bibitem [{\citenamefont {{Reingold}}\ \emph {et~al.}(2019)\citenamefont
  {{Reingold}}, \citenamefont {{Olivas-Gomez}}, \citenamefont {{Simon}},
  \citenamefont {{Arroyo}}, \citenamefont {{Chamberlain}}, \citenamefont
  {{Wurzer}}, \citenamefont {{Spyrou}}, \citenamefont {{Naqvi}}, \citenamefont
  {{Dombos}}, \citenamefont {{Palmisano}}, \citenamefont {{Anderson}},
  \citenamefont {{Clark}}, \citenamefont {{Frentz}}, \citenamefont {{Hall}},
  \citenamefont {{Henderson}}, \citenamefont {{Moylan}}, \citenamefont
  {{Robertson}}, \citenamefont {{Skulski}}, \citenamefont {{Stech}},
  \citenamefont {{Strauss}}, \citenamefont {{Tan}},\ and\ \citenamefont {{Vande
  Kolk}}}]{Craig}%
  \BibitemOpen
  \bibfield  {author} {\bibinfo {author} {\bibfnamefont {C.~S.}\ \bibnamefont
  {{Reingold}}}, \bibinfo {author} {\bibfnamefont {O.}~\bibnamefont
  {{Olivas-Gomez}}}, \bibinfo {author} {\bibfnamefont {A.}~\bibnamefont
  {{Simon}}}, \bibinfo {author} {\bibfnamefont {J.}~\bibnamefont {{Arroyo}}},
  \bibinfo {author} {\bibfnamefont {M.}~\bibnamefont {{Chamberlain}}}, \bibinfo
  {author} {\bibfnamefont {J.}~\bibnamefont {{Wurzer}}}, \bibinfo {author}
  {\bibfnamefont {A.}~\bibnamefont {{Spyrou}}}, \bibinfo {author}
  {\bibfnamefont {F.}~\bibnamefont {{Naqvi}}}, \bibinfo {author} {\bibfnamefont
  {A.~C.}\ \bibnamefont {{Dombos}}}, \bibinfo {author} {\bibfnamefont
  {A.}~\bibnamefont {{Palmisano}}}, \bibinfo {author} {\bibfnamefont
  {T.}~\bibnamefont {{Anderson}}}, \bibinfo {author} {\bibfnamefont {A.~M.}\
  \bibnamefont {{Clark}}}, \bibinfo {author} {\bibfnamefont {B.}~\bibnamefont
  {{Frentz}}}, \bibinfo {author} {\bibfnamefont {M.~R.}\ \bibnamefont
  {{Hall}}}, \bibinfo {author} {\bibfnamefont {S.~L.}\ \bibnamefont
  {{Henderson}}}, \bibinfo {author} {\bibfnamefont {S.}~\bibnamefont
  {{Moylan}}}, \bibinfo {author} {\bibfnamefont {D.}~\bibnamefont
  {{Robertson}}}, \bibinfo {author} {\bibfnamefont {M.}~\bibnamefont
  {{Skulski}}}, \bibinfo {author} {\bibfnamefont {E.}~\bibnamefont {{Stech}}},
  \bibinfo {author} {\bibfnamefont {S.~Y.}\ \bibnamefont {{Strauss}}}, \bibinfo
  {author} {\bibfnamefont {W.~P.}\ \bibnamefont {{Tan}}}, \ and\ \bibinfo
  {author} {\bibfnamefont {B.}~\bibnamefont {{Vande Kolk}}},\ }\href {\doibase
  10.1140/epja/i2019-12748-8} {\bibfield  {journal} {\bibinfo  {journal}
  {European Physical Journal A}\ }\textbf {\bibinfo {volume} {55}},\ \bibinfo
  {eid} {77} (\bibinfo {year} {2019})}\BibitemShut {NoStop}%
\bibitem [{\citenamefont {Olivas-Gomez}\ \emph {et~al.}(2022)\citenamefont
  {Olivas-Gomez}, \citenamefont {Simon}, \citenamefont {Robertson},
  \citenamefont {Dombos}, \citenamefont {Strieder}, \citenamefont {Kadlecek},
  \citenamefont {Hanhardt}, \citenamefont {Kelmar}, \citenamefont {Couder},
  \citenamefont {Görres}, \citenamefont {Stech},\ and\ \citenamefont
  {Wiescher}}]{olivas-gomez_commissioning_2022}%
  \BibitemOpen
  \bibfield  {author} {\bibinfo {author} {\bibfnamefont {O.}~\bibnamefont
  {Olivas-Gomez}}, \bibinfo {author} {\bibfnamefont {A.}~\bibnamefont {Simon}},
  \bibinfo {author} {\bibfnamefont {D.}~\bibnamefont {Robertson}}, \bibinfo
  {author} {\bibfnamefont {A.~C.}\ \bibnamefont {Dombos}}, \bibinfo {author}
  {\bibfnamefont {F.}~\bibnamefont {Strieder}}, \bibinfo {author}
  {\bibfnamefont {T.}~\bibnamefont {Kadlecek}}, \bibinfo {author}
  {\bibfnamefont {M.}~\bibnamefont {Hanhardt}}, \bibinfo {author}
  {\bibfnamefont {R.}~\bibnamefont {Kelmar}}, \bibinfo {author} {\bibfnamefont
  {M.}~\bibnamefont {Couder}}, \bibinfo {author} {\bibfnamefont
  {J.}~\bibnamefont {Görres}}, \bibinfo {author} {\bibfnamefont
  {E.}~\bibnamefont {Stech}}, \ and\ \bibinfo {author} {\bibfnamefont
  {M.}~\bibnamefont {Wiescher}},\ }\href {\doibase
  10.1140/epja/s10050-022-00711-z} {\bibfield  {journal} {\bibinfo  {journal}
  {The European Physical Journal A}\ }\textbf {\bibinfo {volume} {58}},\
  \bibinfo {pages} {57} (\bibinfo {year} {2022})}\BibitemShut {NoStop}%
\bibitem [{\citenamefont {Dombos}\ \emph {et~al.}(2022)\citenamefont {Dombos},
  \citenamefont {Robertson}, \citenamefont {Simon}, \citenamefont {Kadlecek},
  \citenamefont {Hanhardt}, \citenamefont {G\"orres}, \citenamefont {Couder},
  \citenamefont {Kelmar}, \citenamefont {Olivas-Gomez}, \citenamefont {Stech},
  \citenamefont {Strieder},\ and\ \citenamefont {Wiescher}}]{Alex}%
  \BibitemOpen
  \bibfield  {author} {\bibinfo {author} {\bibfnamefont {A.~C.}\ \bibnamefont
  {Dombos}}, \bibinfo {author} {\bibfnamefont {D.}~\bibnamefont {Robertson}},
  \bibinfo {author} {\bibfnamefont {A.}~\bibnamefont {Simon}}, \bibinfo
  {author} {\bibfnamefont {T.}~\bibnamefont {Kadlecek}}, \bibinfo {author}
  {\bibfnamefont {M.}~\bibnamefont {Hanhardt}}, \bibinfo {author}
  {\bibfnamefont {J.}~\bibnamefont {G\"orres}}, \bibinfo {author}
  {\bibfnamefont {M.}~\bibnamefont {Couder}}, \bibinfo {author} {\bibfnamefont
  {R.}~\bibnamefont {Kelmar}}, \bibinfo {author} {\bibfnamefont
  {O.}~\bibnamefont {Olivas-Gomez}}, \bibinfo {author} {\bibfnamefont
  {E.}~\bibnamefont {Stech}}, \bibinfo {author} {\bibfnamefont
  {F.}~\bibnamefont {Strieder}}, \ and\ \bibinfo {author} {\bibfnamefont
  {M.}~\bibnamefont {Wiescher}},\ }\href {\doibase
  10.1103/PhysRevLett.128.162701} {\bibfield  {journal} {\bibinfo  {journal}
  {Phys. Rev. Lett.}\ }\textbf {\bibinfo {volume} {128}},\ \bibinfo {pages}
  {162701} (\bibinfo {year} {2022})}\BibitemShut {NoStop}%
\bibitem [{\citenamefont {Agostinelli}\ \emph {et~al.}(2003)\citenamefont
  {Agostinelli}, \citenamefont {Allison}, \citenamefont {Amako}, \citenamefont
  {Apostolakis}, \citenamefont {Araujo}, \citenamefont {Arce}, \citenamefont
  {Asai}, \citenamefont {Axen}, \citenamefont {Banerjee}, \citenamefont
  {Barrand}, \citenamefont {Behner}, \citenamefont {Bellagamba}, \citenamefont
  {Boudreau}, \citenamefont {Broglia}, \citenamefont {Brunengo}, \citenamefont
  {Burkhardt}, \citenamefont {Chauvie}, \citenamefont {Chuma}, \citenamefont
  {Chytracek}, \citenamefont {Cooperman}, \citenamefont {Cosmo}, \citenamefont
  {Degtyarenko}, \citenamefont {Dell'Acqua}, \citenamefont {Depaola},
  \citenamefont {Dietrich}, \citenamefont {Enami}, \citenamefont {Feliciello},
  \citenamefont {Ferguson}, \citenamefont {Fesefeldt}, \citenamefont {Folger},
  \citenamefont {Foppiano}, \citenamefont {Forti}, \citenamefont {Garelli},
  \citenamefont {Giani}, \citenamefont {Giannitrapani}, \citenamefont {Gibin},
  \citenamefont {{Gómez Cadenas}}, \citenamefont {González}, \citenamefont
  {{Gracia Abril}}, \citenamefont {Greeniaus}, \citenamefont {Greiner},
  \citenamefont {Grichine}, \citenamefont {Grossheim}, \citenamefont
  {Guatelli}, \citenamefont {Gumplinger}, \citenamefont {Hamatsu},
  \citenamefont {Hashimoto}, \citenamefont {Hasui}, \citenamefont {Heikkinen},
  \citenamefont {Howard}, \citenamefont {Ivanchenko}, \citenamefont {Johnson},
  \citenamefont {Jones}, \citenamefont {Kallenbach}, \citenamefont {Kanaya},
  \citenamefont {Kawabata}, \citenamefont {Kawabata}, \citenamefont {Kawaguti},
  \citenamefont {Kelner}, \citenamefont {Kent}, \citenamefont {Kimura},
  \citenamefont {Kodama}, \citenamefont {Kokoulin}, \citenamefont {Kossov},
  \citenamefont {Kurashige}, \citenamefont {Lamanna}, \citenamefont {Lampén},
  \citenamefont {Lara}, \citenamefont {Lefebure}, \citenamefont {Lei},
  \citenamefont {Liendl}, \citenamefont {Lockman}, \citenamefont {Longo},
  \citenamefont {Magni}, \citenamefont {Maire}, \citenamefont {Medernach},
  \citenamefont {Minamimoto}, \citenamefont {{Mora de Freitas}}, \citenamefont
  {Morita}, \citenamefont {Murakami}, \citenamefont {Nagamatu}, \citenamefont
  {Nartallo}, \citenamefont {Nieminen}, \citenamefont {Nishimura},
  \citenamefont {Ohtsubo}, \citenamefont {Okamura}, \citenamefont {O'Neale},
  \citenamefont {Oohata}, \citenamefont {Paech}, \citenamefont {Perl},
  \citenamefont {Pfeiffer}, \citenamefont {Pia}, \citenamefont {Ranjard},
  \citenamefont {Rybin}, \citenamefont {Sadilov}, \citenamefont {{Di Salvo}},
  \citenamefont {Santin}, \citenamefont {Sasaki}, \citenamefont {Savvas},
  \citenamefont {Sawada}, \citenamefont {Scherer}, \citenamefont {Sei},
  \citenamefont {Sirotenko}, \citenamefont {Smith}, \citenamefont {Starkov},
  \citenamefont {Stoecker}, \citenamefont {Sulkimo}, \citenamefont {Takahata},
  \citenamefont {Tanaka}, \citenamefont {Tcherniaev}, \citenamefont {{Safai
  Tehrani}}, \citenamefont {Tropeano}, \citenamefont {Truscott}, \citenamefont
  {Uno}, \citenamefont {Urban}, \citenamefont {Urban}, \citenamefont {Verderi},
  \citenamefont {Walkden}, \citenamefont {Wander}, \citenamefont {Weber},
  \citenamefont {Wellisch}, \citenamefont {Wenaus}, \citenamefont {Williams},
  \citenamefont {Wright}, \citenamefont {Yamada}, \citenamefont {Yoshida},\
  and\ \citenamefont {Zschiesche}}]{geant}%
  \BibitemOpen
  \bibfield  {author} {\bibinfo {author} {\bibfnamefont {S.}~\bibnamefont
  {Agostinelli}}, \bibinfo {author} {\bibfnamefont {J.}~\bibnamefont
  {Allison}}, \bibinfo {author} {\bibfnamefont {K.}~\bibnamefont {Amako}},
  \bibinfo {author} {\bibfnamefont {J.}~\bibnamefont {Apostolakis}}, \bibinfo
  {author} {\bibfnamefont {H.}~\bibnamefont {Araujo}}, \bibinfo {author}
  {\bibfnamefont {P.}~\bibnamefont {Arce}}, \bibinfo {author} {\bibfnamefont
  {M.}~\bibnamefont {Asai}}, \bibinfo {author} {\bibfnamefont {D.}~\bibnamefont
  {Axen}}, \bibinfo {author} {\bibfnamefont {S.}~\bibnamefont {Banerjee}},
  \bibinfo {author} {\bibfnamefont {G.}~\bibnamefont {Barrand}}, \bibinfo
  {author} {\bibfnamefont {F.}~\bibnamefont {Behner}}, \bibinfo {author}
  {\bibfnamefont {L.}~\bibnamefont {Bellagamba}}, \bibinfo {author}
  {\bibfnamefont {J.}~\bibnamefont {Boudreau}}, \bibinfo {author}
  {\bibfnamefont {L.}~\bibnamefont {Broglia}}, \bibinfo {author} {\bibfnamefont
  {A.}~\bibnamefont {Brunengo}}, \bibinfo {author} {\bibfnamefont
  {H.}~\bibnamefont {Burkhardt}}, \bibinfo {author} {\bibfnamefont
  {S.}~\bibnamefont {Chauvie}}, \bibinfo {author} {\bibfnamefont
  {J.}~\bibnamefont {Chuma}}, \bibinfo {author} {\bibfnamefont
  {R.}~\bibnamefont {Chytracek}}, \bibinfo {author} {\bibfnamefont
  {G.}~\bibnamefont {Cooperman}}, \bibinfo {author} {\bibfnamefont
  {G.}~\bibnamefont {Cosmo}}, \bibinfo {author} {\bibfnamefont
  {P.}~\bibnamefont {Degtyarenko}}, \bibinfo {author} {\bibfnamefont
  {A.}~\bibnamefont {Dell'Acqua}}, \bibinfo {author} {\bibfnamefont
  {G.}~\bibnamefont {Depaola}}, \bibinfo {author} {\bibfnamefont
  {D.}~\bibnamefont {Dietrich}}, \bibinfo {author} {\bibfnamefont
  {R.}~\bibnamefont {Enami}}, \bibinfo {author} {\bibfnamefont
  {A.}~\bibnamefont {Feliciello}}, \bibinfo {author} {\bibfnamefont
  {C.}~\bibnamefont {Ferguson}}, \bibinfo {author} {\bibfnamefont
  {H.}~\bibnamefont {Fesefeldt}}, \bibinfo {author} {\bibfnamefont
  {G.}~\bibnamefont {Folger}}, \bibinfo {author} {\bibfnamefont
  {F.}~\bibnamefont {Foppiano}}, \bibinfo {author} {\bibfnamefont
  {A.}~\bibnamefont {Forti}}, \bibinfo {author} {\bibfnamefont
  {S.}~\bibnamefont {Garelli}}, \bibinfo {author} {\bibfnamefont
  {S.}~\bibnamefont {Giani}}, \bibinfo {author} {\bibfnamefont
  {R.}~\bibnamefont {Giannitrapani}}, \bibinfo {author} {\bibfnamefont
  {D.}~\bibnamefont {Gibin}}, \bibinfo {author} {\bibfnamefont
  {J.}~\bibnamefont {{Gómez Cadenas}}}, \bibinfo {author} {\bibfnamefont
  {I.}~\bibnamefont {González}}, \bibinfo {author} {\bibfnamefont
  {G.}~\bibnamefont {{Gracia Abril}}}, \bibinfo {author} {\bibfnamefont
  {G.}~\bibnamefont {Greeniaus}}, \bibinfo {author} {\bibfnamefont
  {W.}~\bibnamefont {Greiner}}, \bibinfo {author} {\bibfnamefont
  {V.}~\bibnamefont {Grichine}}, \bibinfo {author} {\bibfnamefont
  {A.}~\bibnamefont {Grossheim}}, \bibinfo {author} {\bibfnamefont
  {S.}~\bibnamefont {Guatelli}}, \bibinfo {author} {\bibfnamefont
  {P.}~\bibnamefont {Gumplinger}}, \bibinfo {author} {\bibfnamefont
  {R.}~\bibnamefont {Hamatsu}}, \bibinfo {author} {\bibfnamefont
  {K.}~\bibnamefont {Hashimoto}}, \bibinfo {author} {\bibfnamefont
  {H.}~\bibnamefont {Hasui}}, \bibinfo {author} {\bibfnamefont
  {A.}~\bibnamefont {Heikkinen}}, \bibinfo {author} {\bibfnamefont
  {A.}~\bibnamefont {Howard}}, \bibinfo {author} {\bibfnamefont
  {V.}~\bibnamefont {Ivanchenko}}, \bibinfo {author} {\bibfnamefont
  {A.}~\bibnamefont {Johnson}}, \bibinfo {author} {\bibfnamefont
  {F.}~\bibnamefont {Jones}}, \bibinfo {author} {\bibfnamefont
  {J.}~\bibnamefont {Kallenbach}}, \bibinfo {author} {\bibfnamefont
  {N.}~\bibnamefont {Kanaya}}, \bibinfo {author} {\bibfnamefont
  {M.}~\bibnamefont {Kawabata}}, \bibinfo {author} {\bibfnamefont
  {Y.}~\bibnamefont {Kawabata}}, \bibinfo {author} {\bibfnamefont
  {M.}~\bibnamefont {Kawaguti}}, \bibinfo {author} {\bibfnamefont
  {S.}~\bibnamefont {Kelner}}, \bibinfo {author} {\bibfnamefont
  {P.}~\bibnamefont {Kent}}, \bibinfo {author} {\bibfnamefont {A.}~\bibnamefont
  {Kimura}}, \bibinfo {author} {\bibfnamefont {T.}~\bibnamefont {Kodama}},
  \bibinfo {author} {\bibfnamefont {R.}~\bibnamefont {Kokoulin}}, \bibinfo
  {author} {\bibfnamefont {M.}~\bibnamefont {Kossov}}, \bibinfo {author}
  {\bibfnamefont {H.}~\bibnamefont {Kurashige}}, \bibinfo {author}
  {\bibfnamefont {E.}~\bibnamefont {Lamanna}}, \bibinfo {author} {\bibfnamefont
  {T.}~\bibnamefont {Lampén}}, \bibinfo {author} {\bibfnamefont
  {V.}~\bibnamefont {Lara}}, \bibinfo {author} {\bibfnamefont {V.}~\bibnamefont
  {Lefebure}}, \bibinfo {author} {\bibfnamefont {F.}~\bibnamefont {Lei}},
  \bibinfo {author} {\bibfnamefont {M.}~\bibnamefont {Liendl}}, \bibinfo
  {author} {\bibfnamefont {W.}~\bibnamefont {Lockman}}, \bibinfo {author}
  {\bibfnamefont {F.}~\bibnamefont {Longo}}, \bibinfo {author} {\bibfnamefont
  {S.}~\bibnamefont {Magni}}, \bibinfo {author} {\bibfnamefont
  {M.}~\bibnamefont {Maire}}, \bibinfo {author} {\bibfnamefont
  {E.}~\bibnamefont {Medernach}}, \bibinfo {author} {\bibfnamefont
  {K.}~\bibnamefont {Minamimoto}}, \bibinfo {author} {\bibfnamefont
  {P.}~\bibnamefont {{Mora de Freitas}}}, \bibinfo {author} {\bibfnamefont
  {Y.}~\bibnamefont {Morita}}, \bibinfo {author} {\bibfnamefont
  {K.}~\bibnamefont {Murakami}}, \bibinfo {author} {\bibfnamefont
  {M.}~\bibnamefont {Nagamatu}}, \bibinfo {author} {\bibfnamefont
  {R.}~\bibnamefont {Nartallo}}, \bibinfo {author} {\bibfnamefont
  {P.}~\bibnamefont {Nieminen}}, \bibinfo {author} {\bibfnamefont
  {T.}~\bibnamefont {Nishimura}}, \bibinfo {author} {\bibfnamefont
  {K.}~\bibnamefont {Ohtsubo}}, \bibinfo {author} {\bibfnamefont
  {M.}~\bibnamefont {Okamura}}, \bibinfo {author} {\bibfnamefont
  {S.}~\bibnamefont {O'Neale}}, \bibinfo {author} {\bibfnamefont
  {Y.}~\bibnamefont {Oohata}}, \bibinfo {author} {\bibfnamefont
  {K.}~\bibnamefont {Paech}}, \bibinfo {author} {\bibfnamefont
  {J.}~\bibnamefont {Perl}}, \bibinfo {author} {\bibfnamefont {A.}~\bibnamefont
  {Pfeiffer}}, \bibinfo {author} {\bibfnamefont {M.}~\bibnamefont {Pia}},
  \bibinfo {author} {\bibfnamefont {F.}~\bibnamefont {Ranjard}}, \bibinfo
  {author} {\bibfnamefont {A.}~\bibnamefont {Rybin}}, \bibinfo {author}
  {\bibfnamefont {S.}~\bibnamefont {Sadilov}}, \bibinfo {author} {\bibfnamefont
  {E.}~\bibnamefont {{Di Salvo}}}, \bibinfo {author} {\bibfnamefont
  {G.}~\bibnamefont {Santin}}, \bibinfo {author} {\bibfnamefont
  {T.}~\bibnamefont {Sasaki}}, \bibinfo {author} {\bibfnamefont
  {N.}~\bibnamefont {Savvas}}, \bibinfo {author} {\bibfnamefont
  {Y.}~\bibnamefont {Sawada}}, \bibinfo {author} {\bibfnamefont
  {S.}~\bibnamefont {Scherer}}, \bibinfo {author} {\bibfnamefont
  {S.}~\bibnamefont {Sei}}, \bibinfo {author} {\bibfnamefont {V.}~\bibnamefont
  {Sirotenko}}, \bibinfo {author} {\bibfnamefont {D.}~\bibnamefont {Smith}},
  \bibinfo {author} {\bibfnamefont {N.}~\bibnamefont {Starkov}}, \bibinfo
  {author} {\bibfnamefont {H.}~\bibnamefont {Stoecker}}, \bibinfo {author}
  {\bibfnamefont {J.}~\bibnamefont {Sulkimo}}, \bibinfo {author} {\bibfnamefont
  {M.}~\bibnamefont {Takahata}}, \bibinfo {author} {\bibfnamefont
  {S.}~\bibnamefont {Tanaka}}, \bibinfo {author} {\bibfnamefont
  {E.}~\bibnamefont {Tcherniaev}}, \bibinfo {author} {\bibfnamefont
  {E.}~\bibnamefont {{Safai Tehrani}}}, \bibinfo {author} {\bibfnamefont
  {M.}~\bibnamefont {Tropeano}}, \bibinfo {author} {\bibfnamefont
  {P.}~\bibnamefont {Truscott}}, \bibinfo {author} {\bibfnamefont
  {H.}~\bibnamefont {Uno}}, \bibinfo {author} {\bibfnamefont {L.}~\bibnamefont
  {Urban}}, \bibinfo {author} {\bibfnamefont {P.}~\bibnamefont {Urban}},
  \bibinfo {author} {\bibfnamefont {M.}~\bibnamefont {Verderi}}, \bibinfo
  {author} {\bibfnamefont {A.}~\bibnamefont {Walkden}}, \bibinfo {author}
  {\bibfnamefont {W.}~\bibnamefont {Wander}}, \bibinfo {author} {\bibfnamefont
  {H.}~\bibnamefont {Weber}}, \bibinfo {author} {\bibfnamefont
  {J.}~\bibnamefont {Wellisch}}, \bibinfo {author} {\bibfnamefont
  {T.}~\bibnamefont {Wenaus}}, \bibinfo {author} {\bibfnamefont
  {D.}~\bibnamefont {Williams}}, \bibinfo {author} {\bibfnamefont
  {D.}~\bibnamefont {Wright}}, \bibinfo {author} {\bibfnamefont
  {T.}~\bibnamefont {Yamada}}, \bibinfo {author} {\bibfnamefont
  {H.}~\bibnamefont {Yoshida}}, \ and\ \bibinfo {author} {\bibfnamefont
  {D.}~\bibnamefont {Zschiesche}},\ }\href {\doibase
  https://doi.org/10.1016/S0168-9002(03)01368-8} {\bibfield  {journal}
  {\bibinfo  {journal} {Nuclear Instruments and Methods in Physics Research
  Section A: Accelerators, Spectrometers, Detectors and Associated Equipment}\
  }\textbf {\bibinfo {volume} {506}},\ \bibinfo {pages} {250} (\bibinfo {year}
  {2003})}\BibitemShut {NoStop}%
\bibitem [{\citenamefont {Longland}\ \emph {et~al.}(2010)\citenamefont
  {Longland}, \citenamefont {Iliadis}, \citenamefont {Cesaratto}, \citenamefont
  {Champagne}, \citenamefont {Daigle}, \citenamefont {Newton},\ and\
  \citenamefont {Fitzgerald}}]{longland_resonance_2010}%
  \BibitemOpen
  \bibfield  {author} {\bibinfo {author} {\bibfnamefont {R.}~\bibnamefont
  {Longland}}, \bibinfo {author} {\bibfnamefont {C.}~\bibnamefont {Iliadis}},
  \bibinfo {author} {\bibfnamefont {J.~M.}\ \bibnamefont {Cesaratto}}, \bibinfo
  {author} {\bibfnamefont {A.~E.}\ \bibnamefont {Champagne}}, \bibinfo {author}
  {\bibfnamefont {S.}~\bibnamefont {Daigle}}, \bibinfo {author} {\bibfnamefont
  {J.~R.}\ \bibnamefont {Newton}}, \ and\ \bibinfo {author} {\bibfnamefont
  {R.}~\bibnamefont {Fitzgerald}},\ }\href {\doibase
  10.1103/PhysRevC.81.055804} {\bibfield  {journal} {\bibinfo  {journal}
  {Physical Review C}\ }\textbf {\bibinfo {volume} {81}},\ \bibinfo {pages}
  {055804} (\bibinfo {year} {2010})}\BibitemShut {NoStop}%
\bibitem [{\citenamefont {Wang}\ \emph {et~al.}(1991)\citenamefont {Wang},
  \citenamefont {Vogelaar},\ and\ \citenamefont {Kavanagh}}]{11B_Wang}%
  \BibitemOpen
  \bibfield  {author} {\bibinfo {author} {\bibfnamefont {T.~R.}\ \bibnamefont
  {Wang}}, \bibinfo {author} {\bibfnamefont {R.~B.}\ \bibnamefont {Vogelaar}},
  \ and\ \bibinfo {author} {\bibfnamefont {R.~W.}\ \bibnamefont {Kavanagh}},\
  }\href {\doibase 10.1103/PhysRevC.43.883} {\bibfield  {journal} {\bibinfo
  {journal} {Phys. Rev. C}\ }\textbf {\bibinfo {volume} {43}},\ \bibinfo
  {pages} {883} (\bibinfo {year} {1991})}\BibitemShut {NoStop}%
\bibitem [{\citenamefont {Endt}(1990)}]{endt_energy_1990}%
  \BibitemOpen
  \bibfield  {author} {\bibinfo {author} {\bibfnamefont {P.~M.}\ \bibnamefont
  {Endt}},\ }\href {\doibase https://doi.org/10.1016/0375-9474(90)90598-G}
  {\bibfield  {journal} {\bibinfo  {journal} {Nuclear Physics A}\ }\textbf
  {\bibinfo {volume} {521}},\ \bibinfo {pages} {1} (\bibinfo {year}
  {1990})}\BibitemShut {NoStop}%
\bibitem [{\citenamefont {Jaeger}(2001)}]{Jaeger}%
  \BibitemOpen
  \bibfield  {author} {\bibinfo {author} {\bibfnamefont {M.}~\bibnamefont
  {Jaeger}},\ }\href {\doibase 10.18419/opus-4673} {\  (\bibinfo {year}
  {2001}),\ 10.18419/opus-4673}\BibitemShut {NoStop}%
\bibitem [{\citenamefont {Harms}\ \emph {et~al.}(1991)\citenamefont {Harms},
  \citenamefont {Kratz},\ and\ \citenamefont {Wiescher}}]{Harms}%
  \BibitemOpen
  \bibfield  {author} {\bibinfo {author} {\bibfnamefont {V.}~\bibnamefont
  {Harms}}, \bibinfo {author} {\bibfnamefont {K.-L.}\ \bibnamefont {Kratz}}, \
  and\ \bibinfo {author} {\bibfnamefont {M.}~\bibnamefont {Wiescher}},\ }\href
  {\doibase 10.1103/PhysRevC.43.2849} {\bibfield  {journal} {\bibinfo
  {journal} {Phys. Rev. C}\ }\textbf {\bibinfo {volume} {43}},\ \bibinfo
  {pages} {2849} (\bibinfo {year} {1991})}\BibitemShut {NoStop}%
\bibitem [{\citenamefont {Giesen}\ \emph {et~al.}(1993)\citenamefont {Giesen},
  \citenamefont {Browne}, \citenamefont {Görres}, \citenamefont {Graff},
  \citenamefont {Iliadis}, \citenamefont {Trautvetter}, \citenamefont
  {Wiescher}, \citenamefont {Harms}, \citenamefont {Kratz}, \citenamefont
  {Pfeiffer}, \citenamefont {Azuma}, \citenamefont {Buckby},\ and\
  \citenamefont {King}}]{Giesen}%
  \BibitemOpen
  \bibfield  {author} {\bibinfo {author} {\bibfnamefont {U.}~\bibnamefont
  {Giesen}}, \bibinfo {author} {\bibfnamefont {C.~P.}\ \bibnamefont {Browne}},
  \bibinfo {author} {\bibfnamefont {J.}~\bibnamefont {Görres}}, \bibinfo
  {author} {\bibfnamefont {S.}~\bibnamefont {Graff}}, \bibinfo {author}
  {\bibfnamefont {C.}~\bibnamefont {Iliadis}}, \bibinfo {author} {\bibfnamefont
  {H.~P.}\ \bibnamefont {Trautvetter}}, \bibinfo {author} {\bibfnamefont
  {M.}~\bibnamefont {Wiescher}}, \bibinfo {author} {\bibfnamefont
  {W.}~\bibnamefont {Harms}}, \bibinfo {author} {\bibfnamefont {K.~L.}\
  \bibnamefont {Kratz}}, \bibinfo {author} {\bibfnamefont {B.}~\bibnamefont
  {Pfeiffer}}, \bibinfo {author} {\bibfnamefont {R.~E.}\ \bibnamefont {Azuma}},
  \bibinfo {author} {\bibfnamefont {M.}~\bibnamefont {Buckby}}, \ and\ \bibinfo
  {author} {\bibfnamefont {J.~D.}\ \bibnamefont {King}},\ }\href {\doibase
  10.1016/0375-9474(93)90167-V} {\bibfield  {journal} {\bibinfo  {journal}
  {Nuclear Physics A}\ }\textbf {\bibinfo {volume} {561}},\ \bibinfo {pages}
  {95} (\bibinfo {year} {1993})}\BibitemShut {NoStop}%
\bibitem [{\citenamefont {Ota}\ \emph {et~al.}(2020)\citenamefont {Ota},
  \citenamefont {Christian}, \citenamefont {Lotay}, \citenamefont {Catford},
  \citenamefont {Bennett}, \citenamefont {Dede}, \citenamefont {Doherty},
  \citenamefont {Hallam}, \citenamefont {Hooker}, \citenamefont {Hunt},
  \citenamefont {Jayatissa}, \citenamefont {Matta}, \citenamefont {Moukaddam},
  \citenamefont {Rogachev}, \citenamefont {Saastamoinen}, \citenamefont
  {Tostevin}, \citenamefont {Upadhyayula},\ and\ \citenamefont
  {Wilkinson}}]{ota_decay_2020}%
  \BibitemOpen
  \bibfield  {author} {\bibinfo {author} {\bibfnamefont {S.}~\bibnamefont
  {Ota}}, \bibinfo {author} {\bibfnamefont {G.}~\bibnamefont {Christian}},
  \bibinfo {author} {\bibfnamefont {G.}~\bibnamefont {Lotay}}, \bibinfo
  {author} {\bibfnamefont {W.~N.}\ \bibnamefont {Catford}}, \bibinfo {author}
  {\bibfnamefont {E.~A.}\ \bibnamefont {Bennett}}, \bibinfo {author}
  {\bibfnamefont {S.}~\bibnamefont {Dede}}, \bibinfo {author} {\bibfnamefont
  {D.~T.}\ \bibnamefont {Doherty}}, \bibinfo {author} {\bibfnamefont
  {S.}~\bibnamefont {Hallam}}, \bibinfo {author} {\bibfnamefont
  {J.}~\bibnamefont {Hooker}}, \bibinfo {author} {\bibfnamefont
  {C.}~\bibnamefont {Hunt}}, \bibinfo {author} {\bibfnamefont {H.}~\bibnamefont
  {Jayatissa}}, \bibinfo {author} {\bibfnamefont {A.}~\bibnamefont {Matta}},
  \bibinfo {author} {\bibfnamefont {M.}~\bibnamefont {Moukaddam}}, \bibinfo
  {author} {\bibfnamefont {G.~V.}\ \bibnamefont {Rogachev}}, \bibinfo {author}
  {\bibfnamefont {A.}~\bibnamefont {Saastamoinen}}, \bibinfo {author}
  {\bibfnamefont {J.~A.}\ \bibnamefont {Tostevin}}, \bibinfo {author}
  {\bibfnamefont {S.}~\bibnamefont {Upadhyayula}}, \ and\ \bibinfo {author}
  {\bibfnamefont {R.}~\bibnamefont {Wilkinson}},\ }\href {\doibase
  https://doi.org/10.1016/j.physletb.2020.135256} {\bibfield  {journal}
  {\bibinfo  {journal} {Physics Letters B}\ }\textbf {\bibinfo {volume}
  {802}},\ \bibinfo {pages} {135256} (\bibinfo {year} {2020})}\BibitemShut
  {NoStop}%
\bibitem [{\citenamefont {Jayatissa}\ \emph {et~al.}(2020)\citenamefont
  {Jayatissa}, \citenamefont {Rogachev}, \citenamefont {Goldberg},
  \citenamefont {Koshchiy}, \citenamefont {Christian}, \citenamefont {Hooker},
  \citenamefont {Ota}, \citenamefont {Roeder}, \citenamefont {Saastamoinen},
  \citenamefont {Trippella}, \citenamefont {Upadhyayula},\ and\ \citenamefont
  {Uberseder}}]{JAYATISSA}%
  \BibitemOpen
  \bibfield  {author} {\bibinfo {author} {\bibfnamefont {H.}~\bibnamefont
  {Jayatissa}}, \bibinfo {author} {\bibfnamefont {G.}~\bibnamefont {Rogachev}},
  \bibinfo {author} {\bibfnamefont {V.}~\bibnamefont {Goldberg}}, \bibinfo
  {author} {\bibfnamefont {E.}~\bibnamefont {Koshchiy}}, \bibinfo {author}
  {\bibfnamefont {G.}~\bibnamefont {Christian}}, \bibinfo {author}
  {\bibfnamefont {J.}~\bibnamefont {Hooker}}, \bibinfo {author} {\bibfnamefont
  {S.}~\bibnamefont {Ota}}, \bibinfo {author} {\bibfnamefont {B.}~\bibnamefont
  {Roeder}}, \bibinfo {author} {\bibfnamefont {A.}~\bibnamefont
  {Saastamoinen}}, \bibinfo {author} {\bibfnamefont {O.}~\bibnamefont
  {Trippella}}, \bibinfo {author} {\bibfnamefont {S.}~\bibnamefont
  {Upadhyayula}}, \ and\ \bibinfo {author} {\bibfnamefont {E.}~\bibnamefont
  {Uberseder}},\ }\href {\doibase
  https://doi.org/10.1016/j.physletb.2020.135267} {\bibfield  {journal}
  {\bibinfo  {journal} {Physics Letters B}\ }\textbf {\bibinfo {volume}
  {802}},\ \bibinfo {pages} {135267} (\bibinfo {year} {2020})}\BibitemShut
  {NoStop}%
\end{thebibliography}%

\end{document}